\documentclass[pre,aps,floats,twocolumn,superscriptaddress,floatfix]{revtex4}
\usepackage{graphicx,amssymb,amsmath,ifthen,rotating}
\usepackage[active]{srcltx}
\begin{document}
\title{Generalized Pauli principle for particles with distinguishable traits}  
\author{Dan Liu}
\affiliation{
  Department of Physics,
  University of Rhode Island,
  Kingston RI 02881, USA}
  \author{Jared Vanasse}
\affiliation{
  Department of Physics,
  University of Rhode Island,
  Kingston RI 02881, USA}
\author{Gerhard M{\"{u}}ller}
\affiliation{
  Department of Physics,
  University of Rhode Island,
  Kingston RI 02881, USA}
\author{Michael Karbach}
\affiliation{
  Bergische Universit{\"{a}}t Wuppertal,
  Fachgruppe Physik,
  D-42097 Wuppertal, Germany}
\begin{abstract}
  The $s=\frac{3}{2}$ Ising spin chain with uniform nearest-neighbor coupling, quadratic single-site potential,
  and magnetic field is shown to be equivalent to a system of 17 species of particles with internal structure.
  The same set of particles (with different energies) is shown to generate the spectrum 
  of the $s=\frac{1}{2}$ Ising chain with dimerized nearest-neighbor coupling.
  The particles are free of interaction energies even at high densities.
  The mutual exclusion statistics of particles from all species is determined by their internal
  structure and encoded in a generalized Pauli principle.
  The exact statistical mechanical analysis can be performed for thermodynamically open or closed 
  systems and with arbitrary energies assigned to all particle species.
  Special circumstances make it possible to merge two or more species into a single species.
  All traits that distinguish the original species become ignorable.
  The particles from the merged species are effectively indistinguishable and obey modified 
  exclusion statistics.
  Different mergers may yield the same endproduct, implying that the inverse process
  (splitting any species into subspecies) is not unique.
  In a macroscopic system of two merged species at thermal equilibrium, the concentrations of the 
  original species satisfy a functional relation governed by their mutual statistical interaction.
  That relation is derivable from an extremum principle.
  In the Ising context the system is open and the particle energies depend on the Hamiltonian parameters.  
  Simple models of polymerization and solitonic paramagnetism each represent a closed system
  of two species that can transform into each other. 
  Here they represent distinguishable traits with different energies of the same physical particle.
  \end{abstract}

\maketitle

%
\section{Introduction}\label{sec:intro}
%
The idea that matter consists of particles, though ancient, has remained controversial 
until the dawn of the twentieth century.
All known particles are classified into species hierarchically, e.g. at the level of quarks, 
at the level of baryons, mesons, and leptons, at the level of atoms, and at the level of molecules.
The indistinguishability of particles from the same species -- a key concept of statistical mechanics
that naturally arose from symmetry requirements in quantum mechanics --
has clear-cut consequences in the microscopic and macroscopic realms of physical reality 
\cite{Nolt09,Reic98}.
The universality of this concept is supported by overwhelming evidence.

The notion of particle species in the context of indistinguishability is subtle.
Two electrons may be distinguishable by their spin orientation, their momenta, 
or their angular momenta.
Different isotopes distinguish atoms of the same element, and different isomers distinguish 
molecules of the same composition.
All distinguishable traits are observable attributes.
Particles that share all traits are identical.
However, particles need not be identical to belong to the same species.
Not all distinguishable traits are relevant in all contexts.

For the purpose of this study we use \emph{motifs} to characterize distinguishable traits 
of particles.
Particle species with a single motif are classified at the most fundamental level.
At levels of classification that ignore certain traits, particles belonging to one species 
are characterized by one of several motifs.
Inevitably, the particle content (numbers of particles from each species) 
of many-body states is of a different mix at different levels of classification.
However, the statistical mechanical analysis of a system of particles must produce 
consistent results at all levels of classification. 
Understanding these consistency criteria is one of our goals.

Aiming to hide all distinguishable traits that do not affect the physical quantities under investigation
minimizes the number of species and assigns multiple motifs to some or all of them.
It is one optimizing factor for the analysis.
The second optimizing factor selects particle species such as to weaken or eliminate 
interaction energies between particles.
The two operate in the same arena but have different ends and are subject to different constraints.

In two previous papers we established a scheme of classifying statistically interacting particles
with shapes into species according to structure and into categories according to (primitive)
functions.
We investigated the criteria that make the particles free of interaction energies \cite{copic, picnnn}.
In this work we demonstrate how multiple species of statistically interacting particles are merged 
or split according to need and economy in a way that produces consistent and exact results at all 
levels of classification.
We again select a particular many-body system to develop an idea of much wider scope.

We consider the spin-$\frac{3}{2}$ Ising chain with nearest-neighbor coupling, a quadratic single-site 
potential, a magnetic field, and periodic boundary conditions:
\begin{equation}\label{eq:1} 
\mathcal{H}=\sum_{l=1}^N\Big[JS_l^zS_{l+1}^z+D\big[(S_l^z)^2- \tfrac{1}{4}\big]-hS_l^z\Big],
\end{equation}
where $S_l^z=\pm\frac{3}{2}, \pm\frac{1}{2}$.
It is sufficiently complex to facilitate multiple levels of classification for the particles assembled 
by the spin coupling.
For the product eigenstates we use the notation 
$|\sigma_1\sigma_2\cdots\sigma_N\rangle_p$,
where the periodicity $p$ denotes the number of distinct vectors related via translations.
To the site variables $\sigma_l$ we assign symbols as follows: 
$(+\frac{3}{2},+\frac{1}{2},-\frac{1}{2},-\frac{3}{2}) ~\hat{=}~ 
(\Uparrow, \uparrow, \downarrow, \Downarrow)$.

There exist four zero-temperature phases at $h=0$ in sectors of the $(J,D)$-plane:
\begin{align}\label{eq:h0phases}
& \Phi_f:~ |\uparrow\uparrow\uparrow\cdots\rangle_1, 
|\downarrow\downarrow\downarrow\cdots\rangle_1~ \mathrm{at}~ J<0, D>-J; \nonumber \\
& \Phi_F:~ |\Uparrow\Uparrow\Uparrow\cdots\rangle_1,
|\Downarrow\Downarrow\Downarrow\cdots\rangle_1~ \mathrm{at}~ J<0, D<-J; \nonumber \\
& \Phi_a:~ |\uparrow\downarrow\uparrow\cdots\rangle_2~ \mathrm{at}~ J>0, D>J; \nonumber \\
& \Phi_A:~ |\Uparrow\Downarrow\Uparrow\cdots\rangle_2~ \mathrm{at}~ J>0, D<J.
\end{align}

\begin{figure}[htb]
  \centering
\includegraphics[width=78mm]{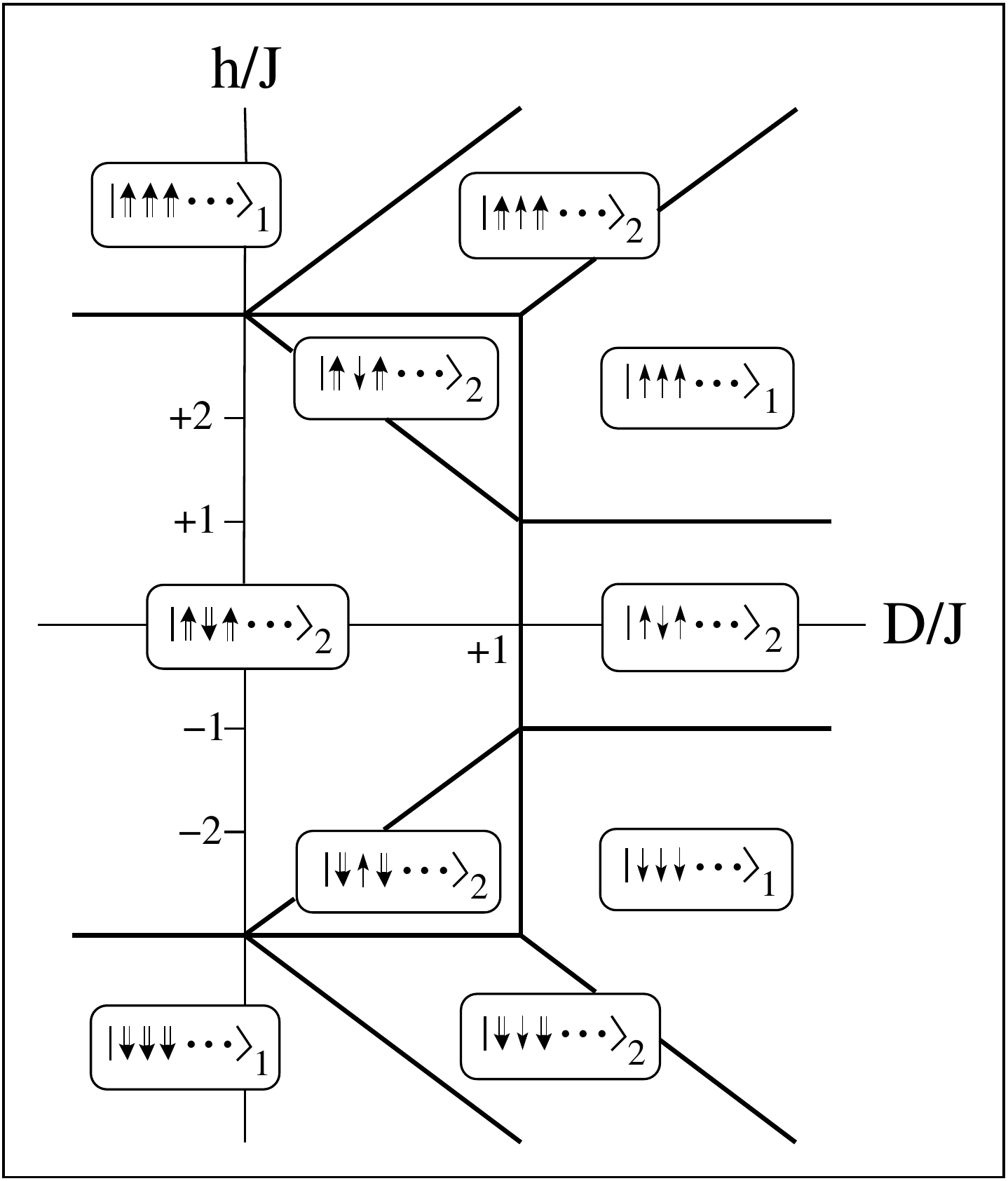} 
  \caption{$T=0$ phases of model system (\ref{eq:1}).} 
  \label{fig:dlj3-pdh}
\end{figure}

At $h\neq0$ the number of phases increases to ten \cite{CJSZ03}.
All of them are realized for $J>0$. Their ranges in the reduced parameter plane $(D/J,h/J)$ 
are shown in Fig.~\ref{fig:dlj3-pdh}.
The magnetization increases from zero to saturation in one step if $D/J<0$ 
and in three steps if $D/J>0$,
The plateau phases are at one third and two thirds of the saturation magnetization.
Four phases are non-degenerate and six are twofold degenerate.
When $h$ increases from zero at $D/J>1$ the translational symmetry increases at the 
first transition, decreases at the second, and increases again at the third.
In the limit $D\to\infty$ the second term in (\ref{eq:1}) reduces the Hilbert space of finite-energy
states to that of the $s=\frac{1}{2}$ Ising spin chain.

In Sec.~\ref{sec:flopa} we introduce, for the $s=\frac{3}{2}$ Ising chain, one set of 17 species of 
particles that remain free of interaction energies even at high densities. 
Two species do the same trick in the limit $D\to\infty$. 
In Sec.~\ref{sec:spms} we formulate criteria for merging two or more species into one
via combinatorial analysis.
In Sec.~\ref{sec:os} we discuss mergers of species in the context of the statistical mechanical
analysis with applications to open systems. 
Closed systems of particles with distinguishable traits are discussed in Sec.~\ref{sec:cs}.
In Sec.~\ref{sec:concl} the major themes are recapitulated and their threads tied together.

%
\section{Floating particles}\label{sec:flopa}
%
All $4^N$ product eigenstates of (\ref{eq:1}), arrays of $\sigma_l$, 
are now interpreted as strings of particles and elements of pseudo-vacuum,
represented by motifs that comprise between two and four consecutive 
site variables in distinctive patterns.
Motifs interlink along the chain in one shared site variable.

As in previous applications to Ising spin chains with $s=\frac{1}{2},1$ \cite{copic, picnnn, LVP+08},
the coupling of the $\sigma_l$ causes the assembly of particles from a set of 
species that are free of interaction energies 
and generate the entire spectrum of (\ref{eq:1}).
The pseudo-vacuum of choice in this work will be the twofold N\'eel state 
$|\Uparrow\Downarrow\Uparrow\cdots\rangle_2$. It coincides with the physical vacuum in left center 
of the parameter space depicted in Fig.~\ref{fig:dlj3-pdh}.

The search for particles that remain free even at high density is facilitated by the taxonomy 
introduced in Ref.~\cite{copic} using familiar and well-tested notions of fractional exclusion statistics
\cite{Hald91a,Khar05}.
For the multiplicity functions $W(\{N_m\})$, which expresses the number of product 
states that contain specific numbers $\{N_m\}$ of particles from all species, we use 
the same template with no need for adjustments:
\begin{subequations}\label{eq:3} 
\begin{align}
& W(\{N_m\})=\frac{n_{pv}N}{N-N^{(\alpha)}} \prod_{m=1}^M
\left( \begin{tabular}{c}
$d_m+N_m-1$ \\ $N_m$
\end{tabular} \right), \label{eq:3a} \\
& d_m=A_m-\sum_{m'=1}^{M}g_{mm'}(N_{m'}-\delta_{mm'}), \label{eq:3b}  \\
& N^{(\alpha)}=\sum_{m=1}^M\alpha_mN_m, 
\label{eq:3c}
\end{align}
\end{subequations}
where $n_{pv}$ is the multiplicity of the pseudo-vacuum, the $A_m$ are capacity constants,
the $\alpha_m$ are size constants, and the $g_{mm'}$ are statistical interaction coefficients.
Relation (\ref{eq:3b}) embodies the generalized Pauli principle as proposed by 
Haldane~\cite{Hald91a}.
The integer $d_m$ counts the number of open slots for particles of species $m$ in the 
presence of $N_{m'}$ particles from any species $m'$.

The absence of any binding energies, which allows us to view the particles as
floating objects, indistinguishable except for species, is reflected in the energy expression,
\begin{equation}\label{eq:4} 
E\big(\{N_m\}\big)=E_{pv}+\sum_{m=1}^M N_m\epsilon_m,
\end{equation}
for product states with particle content $\{N_m\}$ and particle energies 
$\epsilon_m$ measured from the pseudo-vacuum.
The specifications to be used in (\ref{eq:3}) and (\ref{eq:4}) of the $M=17$ species assembled
by the Ising spin coupling in (\ref{eq:1}) are compiled in Table ~\ref{tab:dlj3-specsneel32}.

\begin{widetext}

\begin{table}[htb]
  \caption{Specifications of $M=17$ particles that generate the spectrum of (\ref{eq:1}) from 
  the state $(n_{pv}=2)$, $|\Uparrow\Downarrow\Uparrow\cdots\rangle_2$: motif, species, energy
  (relative to vacuum at $h=0$), spin, capacity constant, size constant, and statistical 
  interaction coefficients. 
  Segments of $\ell$ vacuum bonds $\Uparrow\Downarrow,\Downarrow\Uparrow$ have energy 
  $E=\frac{1}{4}\ell(8D-9J)$.  For $h\neq0$ the entries of $\epsilon_m$ must be amended by 
  $-s_mh$. Species $m=1,2$ are compacts, $m=3,\ldots,14$ hosts, and $m=15,16,17$ 
  hybrids.}\label{tab:dlj3-specsneel32} 
\begin{center}
{\begin{tabular}{cc|cccc}
motif& $m$ & $2\epsilon_{m}$ & $s_{m}$ & $A_{m}$ &$\alpha_m$
\\ \hline \rule[-2mm]{0mm}{6mm}
$\Uparrow\Uparrow$ & $1$ & $9J$ & $+\frac{3}{2}$ & $\frac{N-1}{2}$ & $1$
\\ \rule[-2mm]{0mm}{5mm}
$\Downarrow\Downarrow$ & $2$ & $9J$ & $-\frac{3}{2}$ & $\frac{N-1}{2}$ & $1$
\\ \rule[-2mm]{0mm}{5mm}
$\Uparrow\downarrow\Uparrow$ & $3$ & $6J-4D$ & $+1$ & 
$\frac{N}{2}-1$ & $2$ \\ \rule[-2mm]{0mm}{5mm}
$\Downarrow\uparrow\Downarrow$ & $4$ & $6J-4D$ 
& $-1$ & $\frac{N}{2}-1$ & $2$ \\ \rule[-2mm]{0mm}{5mm}
$\Uparrow\uparrow\Uparrow$ & $5$ & $12J-4D$ & 
$+2$ & $\frac{N}{2}-1$ & $2$ \\ \rule[-2mm]{0mm}{5mm}
$\Downarrow\downarrow\Downarrow$ & $6$ & $12J-4D$ 
& $-2$ & $\frac{N}{2}-1$ & $2$ \\ \rule[-2mm]{0mm}{5mm}
$\Uparrow\uparrow\Downarrow,\Downarrow\uparrow\Uparrow$ & $7$ &
$9J-4D$ & $+\frac{1}{2}$ & $N-2$ & $2$ \\ \rule[-2mm]{0mm}{5mm}
$\Uparrow\downarrow\Downarrow,\Downarrow\downarrow\Uparrow$ & $8$ & 
$9J-4D$ & $-\frac{1}{2}$ & $N-2$ & $2$ \\ \rule[-2mm]{0mm}{5mm}
$\Uparrow\uparrow\downarrow\Uparrow$ & $9$ & 
$13J-8D$ & $+\frac{3}{2}$ & $\frac{N-3}{2}$ & $3$ \\ \rule[-2mm]{0mm}{5mm}
$\Downarrow\uparrow\downarrow\Downarrow$ & $10$ & 
$13J-8D$ & $-\frac{3}{2}$ & $\frac{N-3}{2}$ & $3$ \\ \rule[-2mm]{0mm}{5mm}
$\Uparrow\downarrow\uparrow\Uparrow$ & $11$ & 
$13J-8D$ & $+\frac{3}{2}$ & $\frac{N-3}{2}$ & $3$ \\ \rule[-2mm]{0mm}{5mm}
$\Downarrow\downarrow\uparrow\Downarrow$ & $12$ & 
$13J-8D$ & $-\frac{3}{2}$ & $\frac{N-3}{2}$ & $3$ \\ \rule[-2mm]{0mm}{5mm}
$\Uparrow\uparrow\downarrow\Downarrow,\Downarrow\downarrow\uparrow\Uparrow$ & $13$ & 
$16J-8D$ & $0$ & $N-3$ & $3$ \\ \rule[-2mm]{0mm}{5mm}
$\Uparrow\downarrow\uparrow\Downarrow,\Downarrow\uparrow\downarrow\Uparrow$ & $14$ & 
$10J-8D$ & $0$ & $N-3$ & $3$ \\ \rule[-2mm]{0mm}{5mm}
$\uparrow\uparrow$ & $15$ & $5J-4D$ & $+\frac{1}{2}$ & 
$0$ & $1$ \\ \rule[-2mm]{0mm}{5mm}
$\downarrow\downarrow$ & $16$ & $5J-4D$ & $-\frac{1}{2}$ & 
$0$ & $1$ \\ \rule[-2mm]{0mm}{5mm}
$\uparrow\downarrow\uparrow,\downarrow\uparrow\downarrow$ & $17$ & 
$8J-8D$ & $0$ & $0$ & $2$
\end{tabular}\hspace{9mm}%
\begin{tabular}{r|rrrrrrrrrrrrrrrrr} 
$g_{mm'}$ & $~1$ & $~2$ & $3$ & $4$ & $5$ & $6$ & $7$ & $8$ & $9$ & $10$ & 
$11$ & $12$ & $13$ & $14$ & $~15$ & $~16$ & $17$ \\
\hline \rule[-2mm]{0mm}{6mm}
$1$ & $\frac{1}{2}$ & $\frac{1}{2}$ & $0$ & $1$ 
& $0$ & $1$ & $\frac{1}{2}$ & $\frac{1}{2}$ 
& $\frac{1}{2}$ & $\frac{3}{2}$ & $\frac{1}{2}$ & $\frac{3}{2}$ 
& $1$ & $1$ & $\frac{1}{2}$ & $\frac{1}{2}$ 
& $1$ \\ \rule[-2mm]{0mm}{5mm}
$2$ & $\frac{1}{2}$ & $\frac{1}{2}$ & $1$ & $0$ 
& $1$ & $0$ & $\frac{1}{2}$ & $\frac{1}{2}$ 
& $\frac{3}{2}$ & $\frac{1}{2}$ & $\frac{3}{2}$ & $\frac{1}{2}$ 
& $1$ & $1$ & $\frac{1}{2}$ & $\frac{1}{2}$ 
& $1$ \\ \rule[-2mm]{0mm}{5mm}
$3$ & $\frac{1}{2}$ & $\frac{1}{2}$ & $1$ & $1$ 
& $1$ & $1$ & $\frac{1}{2}$ & $\frac{1}{2}$ 
& $\frac{1}{2}$ & $\frac{3}{2}$ & $\frac{1}{2}$ & $\frac{3}{2}$ 
& $1$ & $1$ & $\frac{1}{2}$ & $\frac{1}{2}$ 
& $1$ \\ \rule[-2mm]{0mm}{5mm}
$4$ & $\frac{1}{2}$ & $\frac{1}{2}$ & $1$ & $1$ 
& $1$ & $1$ & $\frac{1}{2}$ & $\frac{1}{2}$ 
& $\frac{3}{2}$ & $\frac{1}{2}$ & $\frac{3}{2}$ & $\frac{1}{2}$ 
& $1$ & $1$ & $\frac{1}{2}$ & $\frac{1}{2}$ 
& $1$ \\ \rule[-2mm]{0mm}{5mm}
$5$ & $\frac{1}{2}$ & $\frac{1}{2}$ & $0$ & $1$ 
& $1$ & $1$ & $\frac{1}{2}$ & $\frac{1}{2}$ 
& $\frac{1}{2}$ & $\frac{3}{2}$ & $\frac{1}{2}$ & $\frac{3}{2}$ 
& $1$ & $1$ & $\frac{1}{2}$ & $\frac{1}{2}$ 
& $1$ \\ \rule[-2mm]{0mm}{5mm}
$6$ & $\frac{1}{2}$ & $\frac{1}{2}$ & $1$ & $0$ 
& $1$ & $1$ & $\frac{1}{2}$ & $\frac{1}{2}$ 
& $\frac{3}{2}$ & $\frac{1}{2}$ & $\frac{3}{2}$ & $\frac{1}{2}$ 
& $1$ & $1$ & $\frac{1}{2}$ & $\frac{1}{2}$ 
& $1$ \\ \rule[-2mm]{0mm}{5mm}
$7$ & $1$ & $1$ & $2$ & $2$ 
& $2$ & $2$ & $2$ & $1$ 
& $3$ & $3$ & $3$ & $3$ 
& $3$ & $3$ & $1$ & $1$ 
& $2$ \\ \rule[-2mm]{0mm}{5mm}
$8$ & $1$ & $1$ & $2$ & $2$ 
& $2$ & $2$ & $2$ & $2$ 
& $3$ & $3$ & $3$ & $3$ 
& $3$ & $3$ & $1$ & $1$ 
& $2$ \\ \rule[-2mm]{0mm}{5mm}
$9$ & $\frac{1}{2}$ & $\frac{1}{2}$ & $1$ & $1$ 
& $1$ & $1$ & $\frac{1}{2}$ & $\frac{1}{2}$ 
& $\frac{3}{2}$ & $\frac{3}{2}$ & $\frac{1}{2}$ & $\frac{3}{2}$ 
& $1$ & $1$ & $\frac{1}{2}$ & $\frac{1}{2}$ 
& $1$ \\ \rule[-2mm]{0mm}{5mm}
$10$ & $\frac{1}{2}$ & $\frac{1}{2}$ & $1$ & $1$ 
& $1$ & $1$ & $\frac{1}{2}$ & $\frac{1}{2}$ 
& $\frac{3}{2}$ & $\frac{3}{2}$ & $\frac{3}{2}$ & $\frac{1}{2}$ 
& $1$ & $1$ & $\frac{1}{2}$ & $\frac{1}{2}$ 
& $1$ \\ \rule[-2mm]{0mm}{5mm}
$11$ & $\frac{1}{2}$ & $\frac{1}{2}$ & $1$ & $1$ 
& $1$ & $1$ & $\frac{1}{2}$ & $\frac{1}{2}$ 
& $\frac{3}{2}$ & $\frac{3}{2}$ & $\frac{3}{2}$ & $\frac{3}{2}$ 
& $1$ & $1$ & $\frac{1}{2}$ & $\frac{1}{2}$ 
& $1$ \\ \rule[-2mm]{0mm}{5mm}
$12$ & $\frac{1}{2}$ & $\frac{1}{2}$ & $1$ & $1$ 
& $1$ & $1$ & $\frac{1}{2}$ & $\frac{1}{2}$ 
& $\frac{3}{2}$ & $\frac{3}{2}$ & $\frac{3}{2}$ & $\frac{3}{2}$ 
& $1$ & $1$ & $\frac{1}{2}$ & $\frac{1}{2}$ 
& $1$ \\ \rule[-2mm]{0mm}{5mm}
$13$ & $1$ & $1$ & $2$ & $2$ 
& $2$ & $2$ & $1$ & $1$ 
& $3$ & $3$ & $3$ & $3$ 
& $3$ & $2$ & $1$ & $1$ 
& $2$ \\ \rule[-2mm]{0mm}{5mm}
$14$ & $1$ & $1$ & $2$ & $2$ 
& $2$ & $2$ & $1$ & $1$ 
& $3$ & $3$ & $3$ & $3$ 
& $3$ & $3$ & $1$ & $1$ 
& $2$ \\ \rule[-2mm]{0mm}{5mm}
$15$ & $0$ & $0$ & $0$ & $-1$ 
& $-1$ & $0$ & $-1$ & $0$ 
& $-1$ & $-1$ & $-1$ & $-1$ 
& $-1$ & $-1$ & $0$ & $0$ 
& $-1$ \\ \rule[-2mm]{0mm}{5mm}
$16$ & $0$ & $0$ & $-1$ & $0$ 
& $0$ & $-1$ & $0$ & $-1$ 
& $-1$ & $-1$ & $-1$ & $-1$ 
& $-1$ & $-1$ & $0$ & $0$ 
& $-1$ \\ \rule[-2mm]{0mm}{5mm}
$17$ & $0$ & $0$ & $-1$ & $-1$ 
& $-1$ & $-1$ & $-1$ & $-1$ 
& $-1$ & $-1$ & $-1$ & $-1$ 
& $-1$ & $-1$ & $0$ & $0$ 
& $0$ \\ 
\end{tabular}}
\end{center}
\end{table} 

\end{widetext}

To determine which categories of particles are represented 
in this set we recall \cite{copic} that \emph{compacts} and \emph{hosts} float in segments of 
pseudo-vacuum, whereas \emph{tags} are located inside hosts, and \emph{hybrids} are tags 
with hosting capability or tags that facilitate coexistence with other tags inside the same host.
Hence the first two species are compacts, and the last three are hybrids. 
All other species are hosts. 

The hybrids alone generate $2^N$ product states, each one jammed with particles.
The remaining $4^N-2^N$ states -- the vast majority in a macroscopic system -- contain
at least one compact or host or element of pseudo-vacuum.
Four of the hosts are characterized by two motifs each, an attribute encountered previously 
\cite{copic, picnnn, LVP+08}, which does not cause a counting ambiguity because at most one 
of the two motifs can be placed into any particular slot.

We also consider the spectrum of finite-energy states
in the $2^N$-dimensional Hilbert space pertaining to the limit $D\to\infty$.
Two sets of $M=2$ species generate that spectrum from different pseudo-vacua.
The combinatorics of both sets are governed by multiplicity expression (\ref{eq:3})
with specifications as compiled in Tables~\ref{tab:specsol} and \ref{tab:specht}.
One set features two compacts, the other a host and a tag.
Both sets were previously encountered as a special case in a different context \cite{picnnn}.

\begin{table}[h!]
  \caption{Specifications of $M=2$ species of particles excited from from the N{\'e}el state 
  ($n_{pv}=2$) $|\uparrow\downarrow\uparrow\downarrow\cdots\rangle_2$.
  Segments of $\ell$ vacuum elements, $\uparrow\downarrow, 
  \downarrow\uparrow$, have energy $-\ell J/4$. At $h\neq0$ the entries of $\epsilon_m$ must be 
  amended by $-s_mh$.}\label{tab:specsol} 
\begin{center}
\begin{tabular}{ccc|cccc}
motif & cat. & $m$ & $\epsilon_{m}$ & $s_{m}$ & $A_{m}$ & $\alpha_m$
\\ \hline \rule[-2mm]{0mm}{6mm}
$\uparrow\uparrow$ & comp. & $+$ & $\frac{J}{2}$ & $+\frac{1}{2}$ & $\frac{N-1}{2}$ & $1$
\\ \rule[-2mm]{0mm}{5mm}
$\downarrow\downarrow$ & comp. & $-$ & $\frac{J}{2}$ & $-\frac{1}{2}$ & $\frac{N-1}{2}$ & $1$
\end{tabular}\hspace{8mm}
\begin{tabular}{c|rr}
$g_{mm'}$ & $+$ & $-$  \\ \hline \rule[-2mm]{0mm}{6mm}
$+$ & $~~\frac{1}{2}$ & $~~\frac{1}{2}$ \\ \rule[-2mm]{0mm}{5mm}
$-$ & $\frac{1}{2}$ & $\frac{1}{2}$ 
\end{tabular}
\end{center}
\end{table} 

\begin{table}[h!]
  \caption{Specifications of $M=2$ species of particles excited from the spin-polarized state for 
  ($n_{pv}=1$) $|\uparrow\uparrow\cdots\rangle_1$.
  Segments of $\ell$ vacuum elements, $\uparrow\uparrow$, have energy $\ell J/4$. 
  At $h\neq0$ the entries of $\epsilon_m$ must be amended by $-s_mh$.}\label{tab:specht} 
\begin{center}
\begin{tabular}{ccc|cccc}
motif & cat. & $m$ & $\epsilon_{m}$ & $s_{m}$ & $A_{m}$ & $\alpha_m$
\\ \hline \rule[-2mm]{0mm}{6mm}
$\uparrow\downarrow\uparrow$ & host & $H$ & $-J$ & $-1$ & $N-1$ & $1$
\\ \rule[-2mm]{0mm}{5mm}
$\downarrow\downarrow$ & tag & $T$ & $0$ & $-1$ & $0$ & $1$
\end{tabular}\hspace{8mm}
\begin{tabular}{c|rr}
$g_{mm'}$ & $H$ & $T$  \\ \hline \rule[-2mm]{0mm}{6mm}
$H$ & $2$ & $~~1$ \\ \rule[-2mm]{0mm}{5mm}
$T$ & $-1$ & $0$ 
\end{tabular}
\end{center}
\end{table} 

Particles with the same motifs as in Table~\ref{tab:dlj3-specsneel32} but different energies are relevant 
for the $s=\frac{1}{2}$ Ising model with dimerized nearest-neighbor coupling as described in 
Appendix~\ref{sec:dim}.

%
\section{Species merged or split}\label{sec:spms}
%
In the following we investigate the circumstances under which particle species with multiple motifs
can be produced via mergers from particles with a single motif.
It is also possible, albeit not without ambiguity, to split a species with multiple motifs
into multiple one-motif species \cite{Anghel,Angh10}.
The key criterion is that, at each level of description, particles from all species 
can be treated as indistinguishable, either because they are identical (single motif) or because their 
distinguishable traits (multiple motifs) are not relevant in the given context \cite{Wu10}.
We shall see that the criteria are more stringent on the level of combinatorial analysis
(for systems of all sizes) than on the level of statistical mechanical analysis
(for macroscopic systems).

\subsection{Rules}\label{sec:spms-a}
Consider a set of $M$ species with all specifications that go into the multiplicity expression 
(\ref{eq:3}) and the energy expression (\ref{eq:4}) given.
If, say, the first two species $m=1,2$ are to be merged into a new species $m=0$,
this is possible if their specifications satisfy Anghel's rules \cite{Anghel} and are assigned values
as follows:
\begin{subequations}\label{eq:angh}
\begin{align}\label{eq:angha}
& \epsilon_1=\epsilon_2 \doteq\epsilon_0, \\
\label{eq:anghb}
& \alpha_1=\alpha_2 \doteq\alpha_0, \\
\label{eq:anghc}
& A_1+A_2 \doteq A_0, \\
\label{eq:anghd}
& g_{11}+g_{21}=g_{12}+g_{22}\doteq g_{00}, \\
\label{eq:anghe}
& g_{m1}=g_{m2}\doteq g_{m0},\quad m=3,\ldots,M, \\
\label{eq:anghf}
& g_{1m}+g_{2m}\doteq g_{0m},\quad m=3,\ldots,M.
\end{align}
\end{subequations}
Anghel derived these rules in an effort to ensure consistency in the statistical mechanical analysis
of a system at different levels of coarse-graining \cite{Angh10}.
Rules (\ref{eq:anghc}) and (\ref{eq:anghd}) have to be made more stringent for the combinatorial
analysis of systems of any size.
The multiplicity expressions (\ref{eq:3a})  and the energy
expressions (\ref{eq:4}) of the old and new sets of species must satisfy the relations
\begin{align}\label{eq:murel} 
& W(N_0,N_3,\ldots)=\sum_{\{N_1,N_2\}}
W(N_1,N_2,N_3,\ldots), \\
\label{eq:murele} 
& E(N_0,N_3,\ldots)=\sum_{\{N_1,N_2\}}
E(N_1,N_2,N_3,\ldots),
\end{align}
respectively, where the sum is constrained to pairs with $N_1+N_2=N_0$ as implied by the
definition of the merging operation.
Relation (\ref{eq:murele}) only depends on rule (\ref{eq:angha}).

In proving relation (\ref{eq:murel}) we consider all factors of (\ref{eq:3a}) separately.
The prefactors left and right are identical if condition (\ref{eq:anghb}) holds.
All binomial factors with $m\geq3$ are identical if condition (\ref{eq:anghe}) is satisfied.
We are thus left to prove the equation
\begin{align}\label{eq:a-1} 
\left(\begin{array}{c}
 d_0+N_0-1 \\ N_0 
\end{array}\right)
&=\sum_{N_1=0}^{N_0}
\left(\begin{array}{c}
d_1+N_1-1 \\ N_1 
\end{array}\right) \nonumber \\
& \hspace{9mm}\times \left(\begin{array}{c}
d_2+N_0-N_1-1 \\ N_0-N_1 
\end{array}\right)
\end{align}
with
\begin{align}\label{eq:a-2}
d_0+N_0-1 = D_0+g_{00}-1-(g_{00}-1)N_0, 
\end{align}
\begin{align}
\label{eq:a-3}
d_1+N_1-1 &= D_1+g_{11}-1-g_{12}N_0 \nonumber \\ &+(g_{12}-g_{11}+1)N_1, 
\end{align}
\begin{align}\label{eq:a-4}
d_2+N_0-N_1-1 &= D_2+g_{22}-1-g_{21}N_0 \nonumber \\
&+(g_{21}-g_{22}+1)(N_0-N_1), 
\end{align}
\begin{equation}\label{eq:a-5}
D_m = A_m-\sum_{m'=3}^M g_{mm'}N_{m'},\quad m=0,1,2.
\end{equation}
We note that $D_1+D_2=D_0$ if conditions (\ref{eq:anghc}) and (\ref{eq:anghf}) are satisfied.
Next we enforce condition (\ref{eq:anghd}) by means of the parametrization 
\begin{align}\label{eq:a-6}
& g_{11}=\frac{1}{2}g_{00}+u,\quad g_{12}=\frac{1}{2}g_{00}+v, \nonumber \\
& g_{21}=\frac{1}{2}g_{00}-u,\quad g_{22}=\frac{1}{2}g_{00}-v .
\end{align}
We have identified two scenarios in which (\ref{eq:a-1}) is proven to hold.
Both necessitate a tightening of condition (\ref{eq:anghd}) and one requires 
that condition (\ref{eq:anghc}) be tightened as well.

Type 1 mergers require that $u=v+1$, thus eliminating the last term in both 
(\ref{eq:a-3}) and (\ref{eq:a-4}) and reducing (\ref{eq:a-1}) to the identity
\begin{equation}\label{eq:a-7}
\sum_{i=0}^k
\left(\begin{array}{c} m \\ i \end{array}\right)
\left(\begin{array}{c} n \\ k-i \end{array}\right)=
\left(\begin{array}{c} m+n \\ k \end{array}\right).
\end{equation}
Type 2 mergers require that $u=v=0$ and $A_1=A_2$, implying $D_1=D_2$ and
reducing (\ref{eq:a-1}) to the identity
\begin{equation}\label{eq:a-8}
\sum_{i=0}^k
\left(\begin{array}{c} m+i \\ i \end{array}\right)
\left(\begin{array}{c} m+k-i \\ k-i \end{array}\right)=
\left(\begin{array}{c} 2m+k+1 \\ k \end{array}\right).
\end{equation}
The amendments to rules (\ref{eq:angh}) in generic form thus read
\begin{subequations}\label{eq:amend}
\begin{align}\label{eq:amend1}
& g_{11}=g_{12}\pm 1,\quad g_{22}=g_{21}\pm 1\quad (\mathrm{type}~1), \\
\label{eq:amend2}
& g_{11}=g_{12}=g_{21}=g_{22},\quad A_1=A_2\quad (\mathrm{type}~2).
\end{align}
\end{subequations}

Anghel's rules (\ref{eq:angh}) are determinate when species are merged but some are indeterminate --
even with amendments (\ref{eq:amend}) -- without implied assumptions or additional input when 
species are split \cite{Anghel,Angh10,Wu10}.

\subsection{Applications}\label{sec:spms-b}
We first apply Anghel's rules (\ref{eq:angh}) with amendments (\ref{eq:amend}) 
to the pairs of species listed in Tables~\ref{tab:specsol},~\ref{tab:specht}.
The two compacts can be merged if $\epsilon_+=\epsilon_-$,
which is the case at $h=0$ and arbitrary values of $J$.
The same is true for the host and the tag if $\epsilon_H=\epsilon_T$,
which is the case at $J=0$ and arbitrary values of $h$.

Both mergers, the first being of type 2 and the second of type 1, 
produce one species of lattice fermions with specifications $A_0=N-1$, $g_{00}=1$, $\alpha_0=1$,
signalling fermionic statistics.
The distinguishable traits of the original species are erased from the specifications $A_m$, $g_{mm'}$ 
that go into the statistical mechanical analysis.

In both cases it is possible to replace the two motifs of the original species by a single motif
representing floating lattice fermions.
In the merger of host and tag, a (fermionic) particle $(\bullet)$ is assigned to any down-spin 
and a hole $(\circ)$ to any up-spin.
The mapping is one-on-one with a non-degenerate pseudo-vacuum.
In the merger of the two compacts a particle is assigned to any aligned bond and a hole to 
any anti-aligned bond.
The mapping is two-on-one with a twofold pseudo-vacuum \cite{note1}.

If our goal is to investigate the statistical mechanics of the 17 species of particles listed in 
Table~\ref{tab:dlj3-specsneel32} (via methods explained in Sec.~\ref{sec:os}) we can simplify 
the analysis by first merging all species whose distinguishable traits are not relevant in a given context.
We note that five species already have two motifs each, which means that 
some of their distinguishable traits happen to be redundant in the context of 
model system (\ref{eq:1}) \cite{note3}.

For the special case $h=0$ we find that all twelve species with one motif can be sorted into pairs 
to be merged sequentially on the basis of Anghel's rules (\ref{eq:angh}) and amendment (\ref{eq:amend2}).
Not all sequences are permissible but all permissible sequences yield the same results.
One allowed sequence of mergers is the following:
\begin{align}\label{eq:merge1711}
15~\&~16~\to~\bar{15},\quad 1~\&~2~\to~\bar{1},\quad 5 ~\&~ 6 ~ \to~ \bar{5}, \nonumber \\
3~\&~4~\to~\bar{3},\quad 9~\&~10~\to~\bar{9},\quad 11~\&~12 \to~\bar{11}.
\end{align}
Species 5 and 6, for example, do not satisfy (\ref{eq:anghe}) before the two preceding mergers
have been carried out.
The six mergers produce 11 species with two motifs each and specifications as compiled in 
Table~\ref{tab:dlj3-specsneel32m}. 
None of the mergers can be undone on the basis of the specifications
$\epsilon_m, A_m, \alpha_m, g_{mm'}$ from Table~\ref{tab:dlj3-specsneel32m} and the rules
(\ref{eq:angh}), (\ref{eq:amend2}) alone.
At least some of the contents of Table~\ref{tab:dlj3-specsneel32} will have to be worked out
from scratch via combinatorial analysis of the given motifs.

\begin{table}[t]
  \caption{Specifications of $M=11$ particles produced via six mergers from those listed in 
  Table~\ref{tab:dlj3-specsneel32}. At $h=0$ the spin $s_m$ is an ignorable trait.}
  \label{tab:dlj3-specsneel32m} 
\begin{center}
\begin{tabular}{cc|ccc}
motif& $m$ & $2\epsilon_{m}$ & $A_{m}$ &$\alpha_m$
\\ \hline \rule[-2mm]{0mm}{6mm}
$\Uparrow\Uparrow,\Downarrow\Downarrow$ & $\bar{1}$ & $9J$ & $N-1$ & $1$
\\ \rule[-2mm]{0mm}{5mm}
$\Uparrow\downarrow\Uparrow,\Downarrow\uparrow\Downarrow$ & $\bar{3}$ & $6J-4D$ & 
$N-2$ & $2$ \\ \rule[-2mm]{0mm}{5mm}
$\Uparrow\uparrow\Uparrow,\Downarrow\downarrow\Downarrow$ & $\bar{5}$ & $12J-4D$ 
& $N-2$ & $2$ \\ \rule[-2mm]{0mm}{5mm}
$\Uparrow\uparrow\Downarrow,\Downarrow\uparrow\Uparrow$ & $7$ &
$9J-4D$ & $N-2$ & $2$ \\ \rule[-2mm]{0mm}{5mm}
$\Uparrow\downarrow\Downarrow,\Downarrow\downarrow\Uparrow$ & $8$ & 
$9J-4D$ & $N-2$ & $2$ \\ \rule[-2mm]{0mm}{5mm}
$\Uparrow\uparrow\downarrow\Uparrow,\Downarrow\uparrow\downarrow\Downarrow$ 
& $\bar{9}$ & $13J-8D$ & $N-3$ & $3$ \\ \rule[-2mm]{0mm}{5mm}
$\Uparrow\downarrow\uparrow\Uparrow,\Downarrow\downarrow\uparrow\Downarrow$ 
& $\bar{11}$ & $13J-8D$ & $N-3$ & $3$ \\ \rule[-2mm]{0mm}{5mm}
$\Uparrow\uparrow\downarrow\Downarrow,\Downarrow\downarrow\uparrow\Uparrow$ & $13$ & 
$16J-8D$ & $N-3$ & $3$ \\ \rule[-2mm]{0mm}{5mm}
$\Uparrow\downarrow\uparrow\Downarrow,\Downarrow\uparrow\downarrow\Uparrow$ & $14$ & 
$10J-8D$ & $N-3$ & $3$ \\ \rule[-2mm]{0mm}{5mm}
$\uparrow\uparrow,\downarrow\downarrow$ & $\bar{15}$ & $5J-4D$ & 
$0$ & $1$ \\ \rule[-2mm]{0mm}{5mm}
$\uparrow\downarrow\uparrow,\downarrow\uparrow\downarrow$ & $17$ & 
$8J-8D$ & $0$ & $2$
\end{tabular}

\vspace*{5mm}
\begin{tabular}{r|rrrrrrrrrrr} 
$g_{mm'}$ & $~\bar{1}$ & $~\bar{3}$ & $\bar{5}$ & $7$ & $8$ & $\bar{9}$ & $\bar{11}$ 
& $13$ & $14$ & $\bar{15}$ & $17$ \\
\hline \rule[-2mm]{0mm}{6mm}
$\bar{1}$ & $1$ & $1$ & $1$ & $1$ & $1$ & $2$ & $2$ & $2$ & $2$ & $1$ & $2$ \\ \rule[-2mm]{0mm}{5mm}
$\bar{3}$ & $1$ & $2$ & $2$ & $1$ & $1$ & $2$ & $2$ & $2$ & $2$ & $1$ & $2$ \\ \rule[-2mm]{0mm}{5mm}
$\bar{5}$ & $1$ & $1$ & $2$ & $1$ & $1$ & $2$ & $2$ & $2$ & $2$ & $1$ & $2$ \\ \rule[-2mm]{0mm}{5mm}
$7$ & $1$ & $2$ & $2$ & $2$ & $1$ & $3$ & $3$ & $3$ & $3$ & $1$ & $2$ \\ \rule[-2mm]{0mm}{5mm}
$8$ & $1$ & $2$ & $2$ & $2$ & $2$ & $3$ & $3$ & $3$ & $3$ & $1$ & $2$ \\ \rule[-2mm]{0mm}{5mm}
$\bar{9}$ & $1$ & $2$ & $2$ & $1$ & $1$ & $3$ & $2$ & $2$ & $2$ & $1$ & $2$ \\ \rule[-2mm]{0mm}{5mm}
$\bar{11}$ & $1$ & $2$ & $2$ & $1$ & $1$ & $3$ & $3$ & $2$ & $2$ & $1$ & $2$ \\ \rule[-2mm]{0mm}{5mm}
$13$ & $1$ & $2$ & $2$ & $1$ & $1$ & $3$ & $3$ & $3$ & $2$ & $1$ & $2$ \\ \rule[-2mm]{0mm}{5mm}
$14$ & $1$ & $2$ & $2$ & $1$ & $1$ & $3$ & $3$ & $3$ & $3$ & $1$ & $2$ \\ \rule[-2mm]{0mm}{5mm}
$\bar{15}$ & $0$ & $-1$ & $-1$ & $-1$ & $-1$ & $-2$ & $-2$ & $-2$ & $-2$ & $0$ & $-2$ \\ 
\rule[-2mm]{0mm}{5mm}
$17$ & $0$ & $-1$ & $-1$ & $-1$ & $-1$ & $-1$ & $-1$ & $-1$ & $-1$ & $0$ & $0$ 
\end{tabular}
\end{center}
\end{table} 

%
\section{Open systems}\label{sec:os}
%
In the statistical mechanical analysis of the particles listed in 
Tables~\ref{tab:dlj3-specsneel32}-\ref{tab:dlj3-specsneel32m} the grandcanonical ensemble
is the natural choice.
The average populations of all species are controllable via chemical potential.
Macroscopic systems that are closed for individual species or combination of species
can then be constructed by a switch of independent variables on the level of 
thermodynamic functions (Sec.~\ref{sec:cs}).
In the Ising context the system is open and the chemical potentials are absorbed in the 
particle energies. 

\subsection{Method}\label{sec:os-a}
Our starting point is the expression,
\begin{equation}\label{eq:Zgen} 
Z=\sum_{\{N_m\}}W(\{N_m\})\exp\left(-\sum_{m}\frac{\epsilon_mN_m}{k_BT}\right)
\end{equation}
for the grand partition function of a system of statistically interacting particles
with given energies $\epsilon_m$ and given specifications $A_m$, $\alpha_m$, 
$g_{mm'}$, $n_{pv}$ that go into the multiplicity function (\ref{eq:3}).
Wu's analysis of (\ref{eq:Zgen}) for a generic situation \cite{Wu94} produced the general result
\begin{equation}\label{eq:Zwm} 
Z=\prod_{m}\left(\frac{1+w_m}{w_m}\right)^{A_m},
\end{equation}
where the (real, positive) $w_m$ are solutions of
\begin{equation}\label{eq:wmeq} 
\frac{\epsilon_m}{k_BT}=\ln(1+w_m)-\sum_{m'}g_{m'm}\ln\left(\frac{1+w_{m'}}{w_{m'}}\right).
\end{equation}
The average numbers of particles from each species are inferred from
\begin{equation}\label{eq:Nmwm} 
w_m\langle N_m\rangle+\sum_{m'}g_{mm'}\langle N_{m'}\rangle =A_m.
\end{equation}
The entropy derived from (\ref{eq:Zgen}) can be expressed as a function of the 
$\langle N_m\rangle$ alone:
\begin{subequations}\label{eq:snm}
\begin{align}
S =  k_B\sum_m\Big[&\big(\langle N_{m}\rangle+\langle Y_m\rangle\big)
\ln\big(\langle N_m\rangle+\langle Y_m\rangle\big) \nonumber \\
&\hspace{0mm}-\langle N_m\rangle\ln \langle N_m\rangle 
-\langle Y_m\rangle\ln \langle Y_m\rangle\Big],\\
&  Y_m\doteq A_m-\sum_{m'}g_{mm'} N_{m'}.
\end{align}
\end{subequations}

Merging species in the statistical mechanical analysis by erasing distinguishable traits of no relevance
simplifies the calculation without sacrificing rigor.
If the specifications of any two species, say $m=1,2$, obey Anghel's rules (\ref{eq:angh}), 
the consequences are that
\begin{subequations}\label{eq:wmangh1} 
\begin{align}\label{eq:wmangh1a} 
& w_1=w_2\doteq w_0,\\
\label{eq:wmangh1b} 
& \langle N_1\rangle+\langle N_2\rangle=\langle N_0\rangle,
\end{align}
\end{subequations}
and that species $m=0,3,\ldots$ satisfy Eqs.~(\ref{eq:Zwm})-(\ref{eq:Nmwm})
with the modified $A_m$, $g_{mm'}$.

Interestingly, the additional restrictions imposed by (\ref{eq:amend}) are ignorable.
Implementing the merger in the statistical mechanical framework is less restrictive than
in the combinatorial framework.
The former is good for macroscopic systems, the latter for systems of all sizes.
Type 1 and type 2 mergers leave distinctive signatures even in the
statistical mechanical analysis, nevertheless.

The entropy of mixing of two species of particles with shapes does not necessarily have 
its maximum when both species are present in equal numbers or equal concentrations.
This fact is evident when we compare the entropy expressions (\ref{eq:snm}) before 
and after a merger.
Both can be calculated independently from the $A_m, g_{mm'}$ before and after 
any merger carried out in the framework of the combinatorial analysis (Sec.~\ref{sec:flopa}).
Consistency of the two entropy expressions implies a functional relation between 
$\langle N_1\rangle$, $\langle N_2\rangle$ in addition to (\ref{eq:wmangh1b}).
It can be derived from the extremum principle,
\begin{equation}\label{eq:sextr} 
\left.\frac{\partial }{\partial \langle N_1\rangle}
S(\langle N_1\rangle,\langle N_0\rangle-\langle N_1\rangle,\langle N_3\rangle,\ldots)
\right|_{\langle N_0\rangle}=0.
\end{equation}

\subsection{Two species}\label{sec:os-b}
\begin{figure}[b!]
  \begin{center}
  \includegraphics[width=68mm]{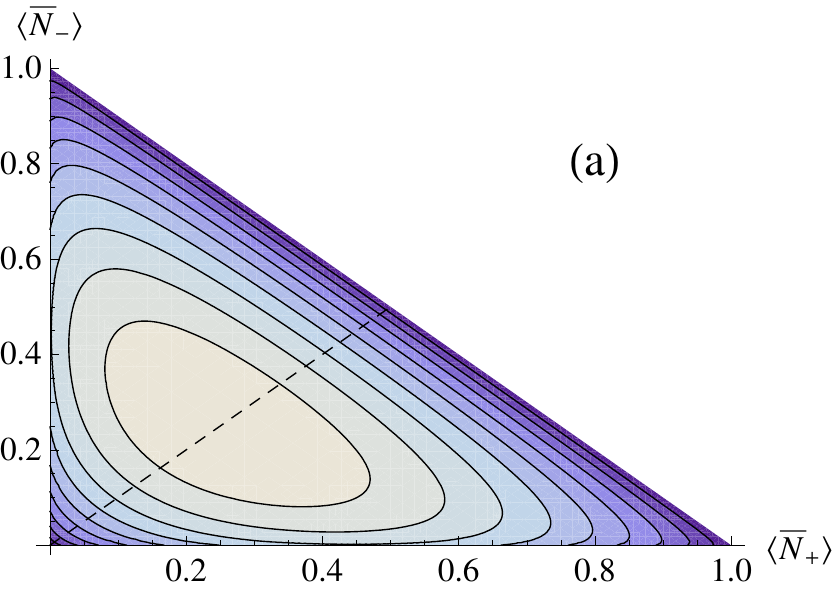}\hspace{3mm}
  \includegraphics[width=68mm]{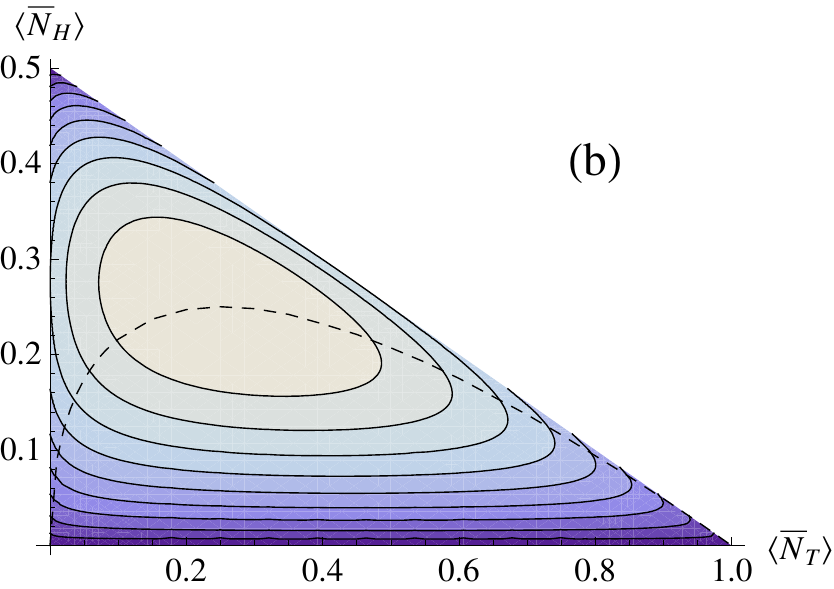}\hspace{3mm}
\end{center}
\caption{Entropy per site, $\bar{S}/k_B$, versus average population densities 
of the two species from  (a) Table~\ref{tab:specsol}
and (b) Table~\ref{tab:specht}.
The contours are at (a) $0.070\ell$, (b) $0.069\ell$, $\ell=1,\ldots,9$.
The dashed lines represent relations (\ref{eq:npmht}).}
  \label{fig:pichs12a}
\end{figure}

In Fig.~\ref{fig:pichs12a} we show contour plots of the entropy landscapes 
$\bar{S}(\langle \bar{N}_+\rangle,\langle \bar{N}_-\rangle)$ and 
$\bar{S}(\langle \bar{N}_H\rangle,\langle \bar{N}_T\rangle)$
pertaining to the two pairs of particle species from Tables~\ref{tab:specsol} and \ref{tab:specht}.
We use the notation $\bar{S}\doteq S/N$ for the entropy per site and 
$\langle\bar{N}_m\rangle\doteq\langle N_m\rangle/N$ for the average population densities.
In generic open systems the $\langle\bar{N}_m\rangle$ are functions of $T,\epsilon_+,\epsilon_-$ 
or $T,\epsilon_H,\epsilon_T$.
In the Ising context, more specifically, the particle energies are functions of the 
Hamiltonian parameters $J,h$.

Both landscapes have triangular shape \cite{note4}.
Panel (a) tells us that each soliton species is able to produce entropy on its own (through
mixing with elements of pseudo-vacuum) and yet higher entropy through mixing with each other at 
moderate densities.
However, when the system is crowded with solitons, the entropy is low no matter what the ratio 
of the two species is. 
Spin-up and spin-down solitons are segregated at the highest density.

From panel (b) we learn that only the hosts can produce entropy on their own, not the tags,
which exist inside hosts in uniform arrays.
Hosts with tags with tags inside come in many different sizes and thus produce yet higher entropy.
The entropy stays high even when the system is close-packed with particles from both species.

Implementing the extremum principle (\ref{eq:sextr}) yields identical entropy expressions,
\begin{equation}\label{eq:sferm} 
\bar{S}/k_B
=-\langle \bar{N}_0\rangle\ln\langle \bar{N}_0\rangle
-\Big(1-\langle \bar{N}_0\rangle\Big)\ln\Big(1-\langle \bar{N}_0\rangle\Big),
\end{equation}
reflecting fermionic exclusion statistics, but different functional relations between
the average population densities of the merged species:
\begin{subequations}\label{eq:npmht}
\begin{align}
\langle \bar{N}_+\rangle &=\langle \bar{N}_-\rangle, \\
\langle \bar{N}_H\rangle &=\sqrt{\langle \bar{N}_T\rangle}-\langle \bar{N}_T\rangle.
\end{align}
\end{subequations}

These relations are independent of the particle energies.
They are solely governed by the shapes of the particles and the way they interlink.
Adding fermions with hidden traits of spin-up and spin down solitons produces a fifty-fifty
mix at all densities.
In strong contrast, fermions with hidden host-tag traits are mostly hosts at low density
and mostly tags at high density.

\subsection{Seventeen species}\label{sec:os-c}
The statistical mechanical analysis of (\ref{eq:1}) at $h\neq0$ cannot take advantage of any mergers.
Coupled Eqs.~(\ref{eq:wmeq}) and (\ref{eq:Nmwm}) must be solved for $m=1,\ldots,17$.
Expression (\ref{eq:Zwm}) for the grand partition function can, nevertheless, be written as a function 
of a single $w_m$, 
\begin{equation}\label{eq:Z17} 
Z=\left[(1+w_1)e^{-18K_J+6H}\right]^N,
\end{equation}
where $w_1$ along with $w_2,\ldots,w_{17}$ are functions of  $K_J\doteq J/4k_BT$, 
$K_D\doteq D/4k_BT$, $H\doteq h/2k_BT$.

\begin{figure}[h]
  \begin{center}
  \includegraphics[width=70mm]{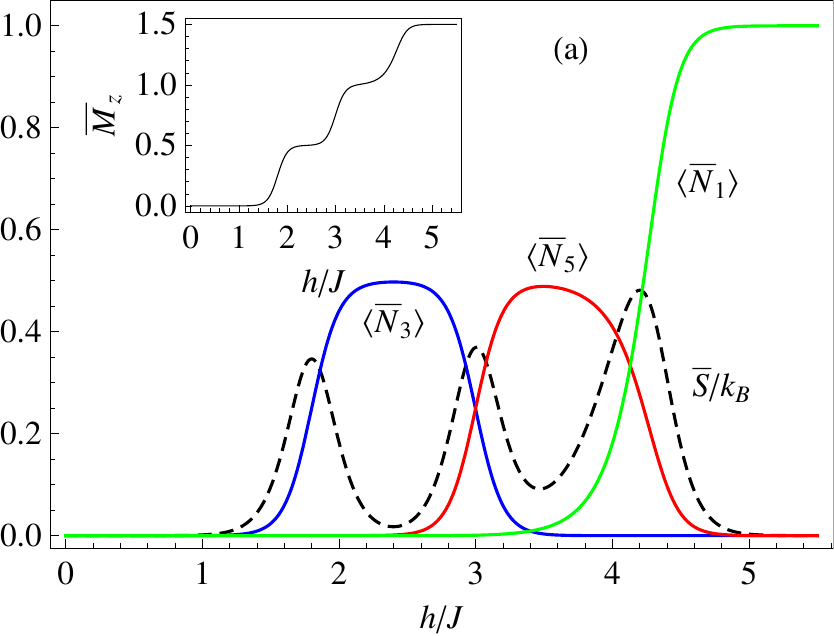}\hspace{3mm}
  \includegraphics[width=70mm]{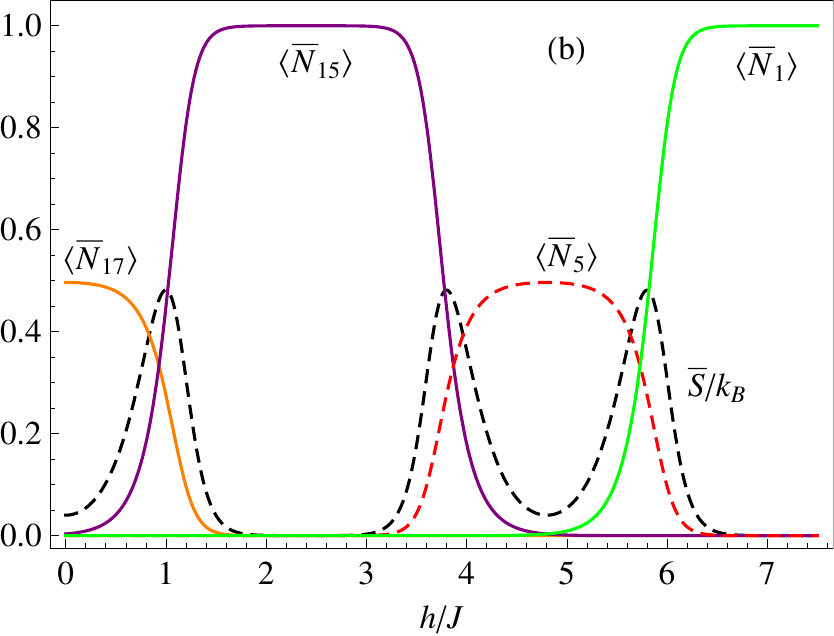}\hspace{3mm}
\end{center}
\caption{Entropy $\bar{S}/k_B$ per site and average population
densities $\langle\bar{N}_{m}\rangle$ for selected species versus $h/J$ at $k_BT/J=0.1$ 
and (a) $D/J=0.6$, (b) $D/J=1.4$. The inset shows the magnetization $\bar{M}_z$ per site
versus $h/J$ for case (a). }
  \label{fig:pichs11}
\end{figure}

In Fig.~\ref{fig:pichs11} we show the dependence of various expectation values on the 
magnetic field along two parallel paths in the parameter space of Fig.~\ref{fig:dlj3-pdh}.
The average population densities $\langle \bar{N}_m\rangle$ inferred from (\ref{eq:wmeq})
and (\ref{eq:Nmwm}) determine the magnetization via
\begin{equation}\label{eq:mzbar} 
\bar{M}_z=\sum_{m} s_m\langle\bar{N}_m\rangle
\end{equation}
and the entropy via (\ref{eq:snm}).

In each case there are two magnetization plateaux, softened by thermal fluctuations (shown for one case).
The entropy shoots up at all three phase boundaries.
Only the third phase boundary is common to both cases.
In each phase region except one the population density of one species is high.
Here the ground state is a solid of that species of particles.
In one phase region, the ground state is the pseudo-vacuum, which, by definition, is devoid of particles.

The ground-state degeneracy is huge at all phase boundaries crossed by the two paths.
Particles or elements of pseudo-vacuum with high densities in adjacent phase regions behave like 
miscible liquids with each spike in the dashed lines representing an entropy of mixing as discussed 
in Ref.~\cite{copic} for a different model.

Many more ground states than the ten phases shown in Fig.~\ref{fig:dlj3-pdh} can be stabilized by 
assigning different energies $\epsilon_m$ to the particles in Table~\ref{tab:dlj3-specsneel32}.
This includes ground states that represent solids of more than one species in ordered or disordered 
configurations.

If we assign the lowest energy (per bond) to particle $m=7$, for example we produce a ferrimagnetic 
state with periodicity $p=4$: $|\Uparrow\uparrow\Downarrow\uparrow\cdots\rangle_4$.
If we assign the same lowest energy per bond also to particle $m=8$ we produce a ground state that is 
antiferrromagnetic but has a Langevin paramagnet embedded in it:
$|\Uparrow\sigma\Downarrow\sigma\cdots\rangle$ with $\sigma=\uparrow,\downarrow$ 
in a random sequence.
For each scenario thus designed the statistical mechanical analysis can be performed exactly.

\subsection{Eleven species}\label{sec:os-d}
At $h=0$ we can take advantage of two layers of simplifications in the analysis.
In Sec.~\ref{sec:spms} we have merged six pairs of the $M=17$ species 
(Table~\ref{tab:dlj3-specsneel32}), thus reducing that number to $M=11$ 
(Table~\ref{tab:dlj3-specsneel32m}), each with two motifs.
The simplifications do not stop here.

Inspection of the coefficients $g_{mm'}$ in Table~\ref{tab:dlj3-specsneel32m} shows that five pairs 
of species almost qualify for further mergers.
The pairs $7~\&~8$, $\bar{9}~\&~\bar{11}$ only violate condition (\ref{eq:anghd}), while pairs
$\bar{3}~\&~\bar{5}$, $\bar{11}~\&~13$, $13~\&~14$ violate also condition (\ref{eq:angha}).
Even though these five pairs of species with $g_{mm}=g_{mm'}=g_{m'm'}=g_{m'm}+1$,
$\epsilon_m-\epsilon_{m'}\doteq\Delta_{mm'}$ cannot be merged, the intact conditions
from (\ref{eq:angh}) lead to further simplifications in Eqs.~(\ref{eq:wmeq}) and 
(\ref{eq:Nmwm}) for those pairs:
\begin{equation}\label{eq:wmangh2} 
\frac{1+w_m}{w_{m'}}=\frac{\langle N_{m'}\rangle}{\langle N_m\rangle}
=\exp\left(\frac{\Delta_{mm'}}{k_BT}\right).
\end{equation}

With all these simplifications in place, Eqs.~(\ref{eq:wmeq}) in exponential form can be rendered 
more compactly:
\begin{align}\label{eq:33-1}
& \frac{1+w_{\bar{1}}}{1+w_{\bar{15}}}=e^{8K_J+8K_D},\quad \frac{w_{\bar{15}}^2}
{1+w_{17}}=e^{4K_J},
\nonumber \\ 
& \frac{w_8w_{\bar{15}}}{w_{17}}=e^{12K_J},\quad 
\frac{w_{\bar{1}}w_8}{1+w_{\bar{5}}}=e^{12K_J},\quad 
\frac{w_{\bar{5}}}{1+w_{\bar{3}}}=e^{12K_J}, \nonumber \\ 
& \frac{1+w_8}{w_7}=1,\quad \frac{w_8w_{14}}{(1+w_7)w_{17}}=e^{4K_J},\quad 
\frac{w_{13}}{1+w_{14}}=e^{12K_J}, \nonumber \\
& \frac{1+w_{\bar{11}}}{w_{13}}=e^{6K_J},\quad \frac{1+w_{\bar{11}}}{w_{\bar{9}}}=1, \nonumber \\
& \frac{(1+w_7)(1+w_{\bar{9}})}{w_{\bar{3}}w_{14}}=e^{12K_J},
\end{align}
and the grand partition function again reduces to (\ref{eq:Z17}), now for $w_1=w_{\bar{1}}$
and $H=0$. 

\begin{figure}[b]
  \begin{center}
 \includegraphics[width=75mm]{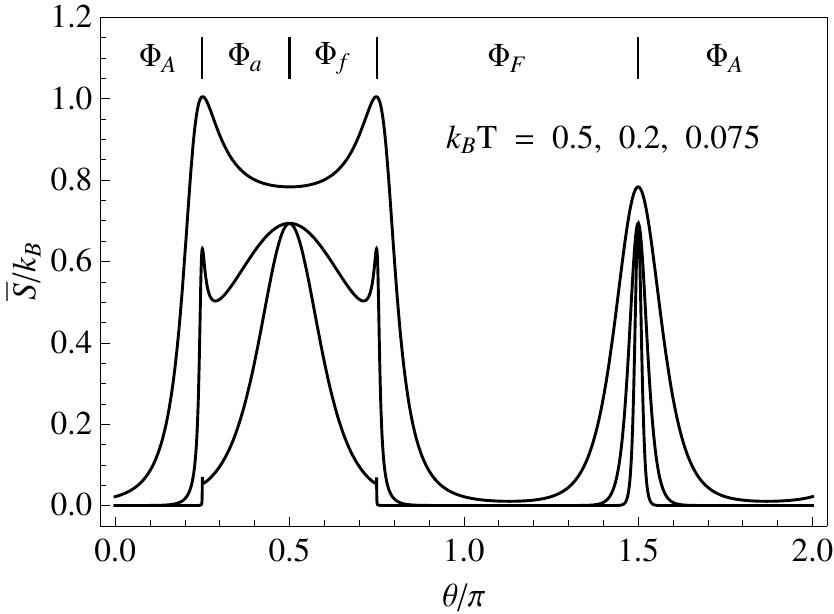}
\end{center}
\caption{Entropy $\bar{S}/k_B$ per site along a circular path, ${J=\cos\theta}$, 
$D=\sin\theta$ in the $(J,D)$ plane at various values of $k_BT$ in arbitrary energy units.
The boundaries between the $T=0$ phases (\ref{eq:h0phases}) are marked by vertical bars.}
  \label{fig:pichs-7a}
\end{figure}

For the discussion of our results we consider a circular path of unit radius in the $(J,D)$ plane. 
In Fig.~\ref{fig:pichs-7a} we show the variation of the entropy (\ref{eq:snm}) along that path
at various various temperatures.
the same arbitrary units are used for $J$, $D$, and $k_BT$.
Naturally, this quantity is sensitive any border crossings of the path between the $T=0$
phases (\ref{eq:h0phases}). 
However, that sensitivity has some unusual features.

The ground-state degeneracy is large at the $\Phi_a-\Phi_f$ and $\Phi_A-\Phi_F$ boundaries
but small at the $\Phi_a-\Phi_A$ and $\Phi_f-\Phi_F$ boundaries.
At $k_BT=0.5$ the entropy $\bar{S}$ has smooth maxima 
of moderate width at three borders and a broad minimum at the fourth.
Surprisingy, an entropy minimum is located at the $\Phi_a-\Phi_f$  border associated with 
a highly degenerate ground state.

Upon lowering the temperature, these extrema evolve in different ways.
The entropy maximum at the $\Phi_A-\Phi_F$ border remains high and sharpens.
The entropy minimum at the $\Phi_a-\Phi_f$ border turns into a maximum of equal height 
but of larger width. 
The maxima at the $\Phi_A-\Phi_a$ and $\Phi_f-\Phi_F$ borders shrink quickly.

\begin{figure}[t]
  \begin{center}
 \includegraphics[width=70mm]{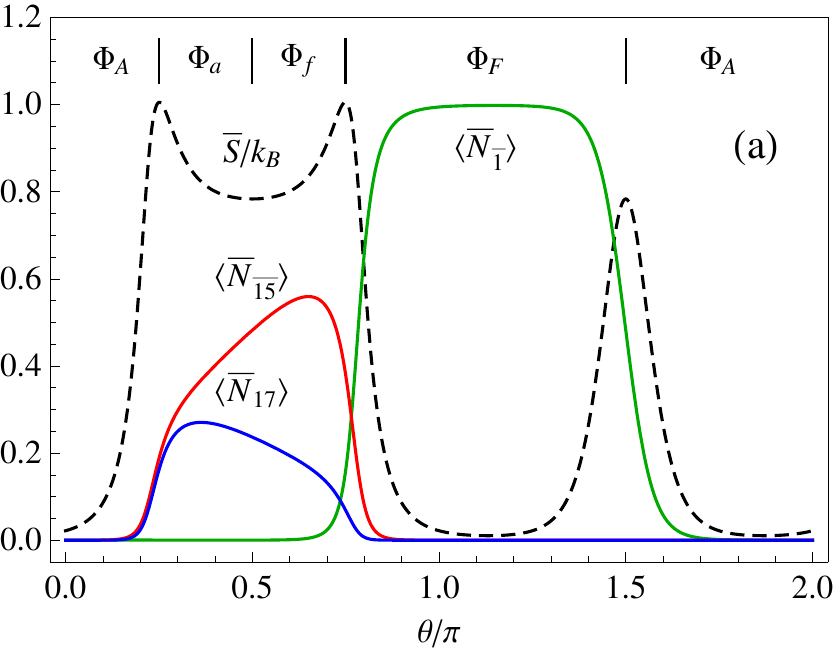}
 \includegraphics[width=70mm]{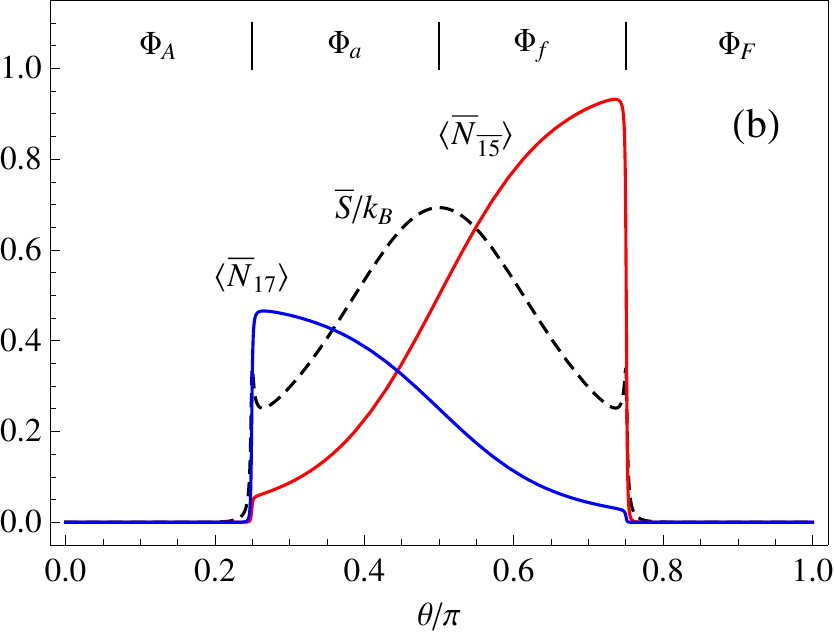}
\end{center}
\caption{Entropy $\bar{S}/k_B$ per site and population densities $\langle\bar{N}_m\rangle$ 
along a circular path, ${J=\cos\theta}$, $D=\sin\theta$, (a) at $k_BT=0.5$ (full circle)
and (b) at $k_BT=0.125$ (half circle) in arbitrary energy units.}
  \label{fig:pichs-7bc}
\end{figure}

Only three of the eleven species, the compact $1$ and the two hybrids $\bar{15}$ and $17$,
are present at high densities in the $(J,D)$ plane.
Their variation along the circular path is shown in Fig.~\ref{fig:pichs-7bc} in
combination with entropy data.

The behavior near the $\Phi_F-\Phi_A$ border is of the kind already discussed in the context of 
Fig.~\ref{fig:pichs11}: compacts $\bar{1}$ and elements of pseudo-vacuum act as miscible liquids.
The contrasting behavior of the data near the $\Phi_A-\Phi_a$ and $\Phi_f-\Phi_F$ borders
can be explained by the observations (discussed in Ref.~\cite{copic} for a different model)
that hybrids $\bar{17}$ and elements of pseudo-vacuum or hybrids $\bar{15}$ and compacts 
$\bar{1}$ act as immiscible liquids.
The population densities vary abruptly and the entropy of mixing is very small at low $T$.

What is unusual about the behavior near the $\Phi_a-\Phi_f$ border is that the energies 
of hybrids $\bar{15}$ and hybrids $\bar{17}$ happen to cross at a very shallow angle.
Both species maintain high population densities across both phases except at very low $T$.
The mixing is not significantly enhanced near the border at moderate $T$.
Hence the absence of a spiked entropy.

%
\section{Closed systems}\label{sec:cs}
%
Once we have identified a set of statistically interacting particles, we are not only free to change 
the energies of some or all species to produce new kinds of ordering as discussed in 
Sec.~\ref{sec:os-c}, we can also impose constraints on the number of particles for individual 
species or combination of species.
Here we present two simple applications that use the particles  from Tables~\ref{tab:specsol}
and \ref{tab:specht} in a closed system.
Both situations illustrate how effects typically attributed to interactions between particles
can be attributed to structures of particles.

\subsection{Soliton paramagnetism}\label{sec:cs-a}
We consider a system of $N_S$ solitons that satisfy semionic exclusion statistics 
(Table~\ref{tab:specsol}). 
The solitons populate a lattice of $N$ sites and are able to switch spin orientation 
through thermal fluctuations.
The solitons have spin $s_\pm=\pm\frac{1}{2}$, therefore, energy $\epsilon_\pm=\mp\frac{1}{2}h$ 
in a magnetic field $h$.

Unlike in the ideal Langevin paramagnet, where the spins are fixed to 
lattice sites and thus distinguishable, the solitons are indistinguishable floating objects.
When an individual soliton flips its spin it must shift its position left or right by one lattice unit:
\begin{equation}\label{eq:solitlr}
|\cdots\uparrow\downarrow\!\bar{\uparrow\uparrow}\!\downarrow\uparrow\cdots\rangle
\to\left\{
\begin{array}{c}
|\cdots\uparrow\!\bar{\downarrow\downarrow}\!\uparrow\downarrow\uparrow\cdots\rangle \\
|\cdots\uparrow\downarrow\uparrow\!\bar{\downarrow\downarrow}\!\uparrow\cdots\rangle
\end{array} \right..
\end{equation}
In crowded conditions only collective spin-flips of groups of solitons are possible.

\begin{figure}[b]
  \begin{center}
 \includegraphics[width=75mm]{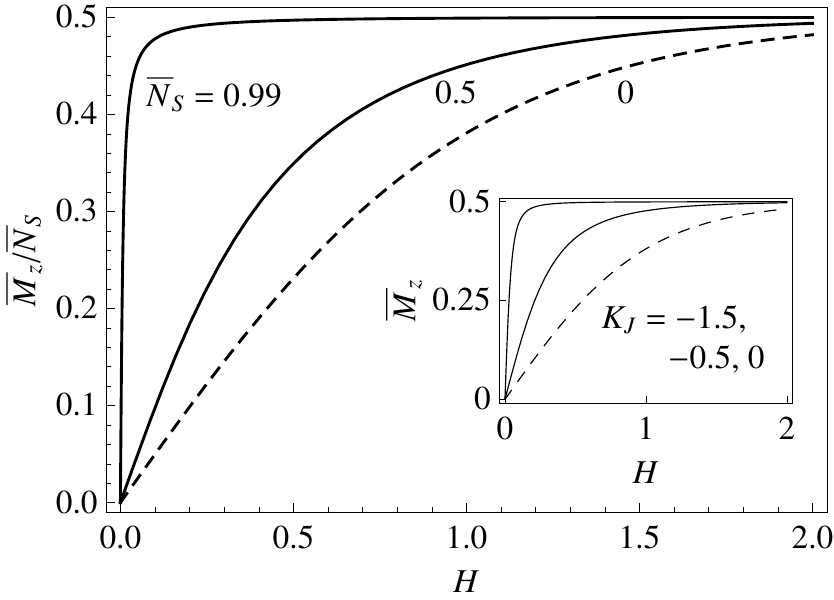}
\end{center}
\caption{Magnetization curves of the soliton paramagnet at various soliton densities
and of the Ising paramagnet at various coupling strengths.}
  \label{fig:pichs-8a}
\end{figure}

We carry out the analysis in the grandcanonical ensemble and enforce a constant (average) 
number of particles, independent of temperature and magnetic field, via a fugacity $\zeta$.
Equations (\ref{eq:wmeq}) in exponentiated form read
\begin{equation}\label{eq:c-2} 
\zeta^{-1}e^{\mp H}=\frac{(1+w_\pm)^{1/2}w_\pm^{1/2}w_\mp^{1/2}}{(1+w_\mp)^{1/2}},
\quad H=\frac{h}{2k_BT},
\end{equation}
yielding the physical solutions
\begin{equation}\label{eq:c-3} 
w_\pm= e^{\mp H}\left[\sqrt{\sinh^2H+\zeta^{-2}}\mp\sinh H\right].
\end{equation}
From Eqs.~(\ref{eq:Nmwm}) we infer the following expressions for the total number of solitons 
per site and the magnetization per site:
\begin{equation}\label{eq:c-4} 
\bar{N}_S\doteq \frac{\langle N_+\rangle+\langle N_-\rangle}{N}
=\frac{q(\zeta,H)\cosh H +\sinh^2H}
{q(\zeta,H)[q(\zeta,H)+\cosh H]},
\end{equation}
\begin{equation}\label{eq:c-5} 
\bar{M}_z\doteq \frac{\langle N_+\rangle-\langle N_-\rangle}{2N}
=\frac{\sinh H}{2q(\zeta,H)},
\end{equation}
where $q(\zeta,H)\doteq \sqrt{\sinh^2H+\zeta^{-2}}$, 
and the following explicit result for the magnetization curves
describing soliton paramagnetism:
\begin{equation}\label{eq:c-6} 
\tilde{M}_z^S\doteq\frac{\bar{M}_z}{\bar{N}_S}=
\frac{1}{2}\frac{Q(\bar{N}_S,H)+\cosh H}
{ Q(\bar{N}_S,H)\coth H+\sinh H},
\end{equation}
where
\begin{align}\label{eq:Qnsh}
Q(\bar{N}_S,H)=\frac{\cosh H}{2\bar{N}_S}&\Big[1-\bar{N}_S \nonumber \\
&\hspace{-10mm}+\sqrt{(1-\bar{N}_S)^2+4\bar{N}_S\tanh^2H}\Big].
\end{align}
Magnetization curves at different soliton densities are shown in Fig.~\ref{fig:pichs-8a}.

Corresponding magnetization curves for the $s=\frac{1}{2}$ Ising paramagnet, Hamiltonian (\ref{eq:1})
in the limit $D\to\infty$ previously considered as an open system of solitons, are described by 
(\ref{eq:c-5}) with $\zeta^{-1}=e^{2K_J}$, $K_J\doteq J/4k_BT$ and have similar shapes.
The common (weak-coupling) Langevin limit, $\tilde{M}_z=\frac{1}{2}\tanh H$, is realized for
$\bar{N}_S, K_J\to0$ and the common strong-coupling limit, 
$\tilde{M}_z=\frac{1}{2}\mathrm{sgn}H$, for $\bar{N}_S\to1$, $K_J\to\infty$.

Away from the Langevin limit (dashed curves), the magnetization is enhanced due to suppressed
thermal fluctuations. 
However, different physical mechanisms are responsible in the two cases.
In the Ising paramagnet individual spin flips are suppressed by a dynamic effect (energy barrier).
In the soliton paramagnet, individual spin flips are suppressed by a kinematic constraint 
(space limitation).

\subsection{Polymerization}\label{sec:cs-b}
Consider a system of $N_M$ molecules from a single species with a propensity for linking up into 
polymers.
The molecules inhabit a one-dimensional lattice of $N$ sites.
Individual molecules have a certain size and give up a fraction of that size with each link to a 
neighbor.
Associated with each link is a binding energy $\epsilon_B$.
Positive (negative) values enhance (suppress) polymerization.
Crowding may drive polymerization even for negative binding energies.

What fraction of molecules are bound in polymers and what is the average length of a polymer 
at any given temperature?
With the host and tag particles from Table~\ref{tab:specht} we can answer these questions 
in purely kinematic terms.

Hosts $(\uparrow\downarrow\uparrow)$ are monomers or seeds of polymers.
Tags $(\downarrow\downarrow)$ are interior molecules of polymers.
Each polymer thus consists of one host and at least one tag
$(\uparrow\downarrow\uparrow+\downarrow\downarrow
=\uparrow\downarrow\downarrow\uparrow)$.
The number of bonds in a polymer is equal to the number of tags it contains.
The centers of two nearest-neighbor monomers are two lattice units apart.
That distance is reduced to one lattice unit when monomers form a dimer \cite{note5}.

The binding energy $\epsilon_B$ is accounted for by assigning different energies to hosts and tags:
$\epsilon_B=\epsilon_H-\epsilon_T$.
Equations~(\ref{eq:wmeq}) for this situation become
\begin{equation}\label{eq:weqpol} 
\frac{1}{\zeta}=\frac{w_H^2(1+w_T)}{w_T(1+w_H)},\quad 
\frac{e^{-K_B}}{\zeta}=\frac{w_H(1+w_T)}{1+w_H},
\end{equation}
where $K_B=\epsilon_B/k_BT$.
We have absorbed the host energy $\epsilon_H$ in the fugacity $\zeta$.
The physically relevant solutions for an open system are
\begin{align}\label{eq:wmsolpol} 
w_H=w_Te^{K_B}=\frac{1}{2\zeta}\Big[1-\zeta e^{K_B} +\sqrt{(1-\zeta e^{K_B})^2+4\zeta}\Big].
\end{align}

The population densities $\langle\bar{N}_m\rangle\doteq\langle N_m\rangle/N$, $m=H,T$,
inferred from Eqs.~(\ref{eq:Nmwm}) are
\begin{equation}\label{eq:NHT} 
\langle\bar{N}_H\rangle=w_T\langle\bar{N}_T\rangle=
\frac{w_T}{w_T(w_H+2)+1}.
\end{equation}
Now we switch to a closed system by imposing the constraint
$\langle\bar{N}_H\rangle+\langle\bar{N}_T\rangle=\bar{N}_M$.
Replacing $\zeta$ by $\bar{N}_M$ as independent variable we write
\begin{align}\label{eq:zetatonm} 
w_H&=\frac{1}{2\bar{N}_M}\Big[1-2\bar{N}_M \nonumber \\
&+\sqrt{(1-2\bar{N}_M)^2+4\bar{N}_M(1-\bar{N}_M)e^{K_B}}\Big].
\end{align}

\begin{figure}[t]
  \begin{center}
 \includegraphics[width=75mm]{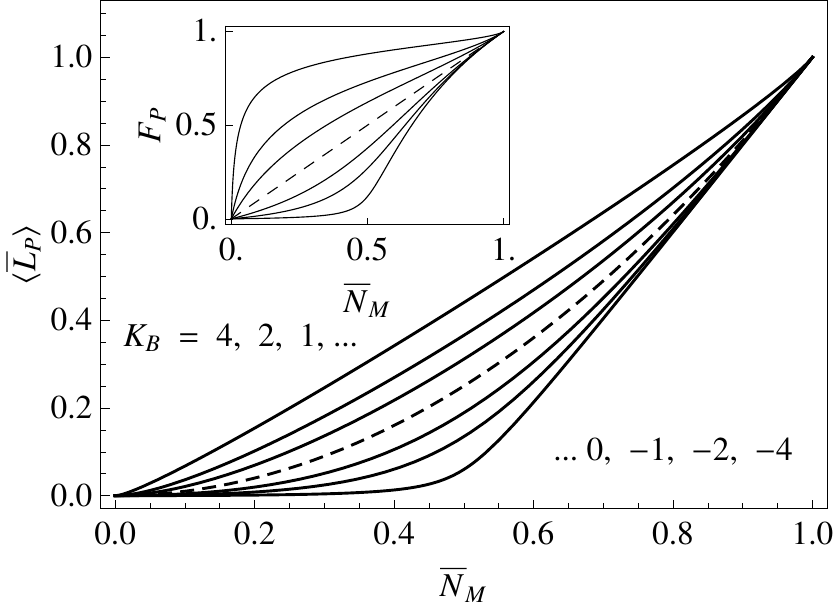}
\end{center}
\caption{Average length of polymer (main plot) and fraction of molecules that are bound in a polymer
(inset) versus molecular population density at various values of $\epsilon_B/k_BT$. The dashed lines
are for $\epsilon_B=0$.}
  \label{fig:pichs-9a}
\end{figure}

In Fig.~\ref{fig:pichs-9a} we show the dependence on the molecular density $\bar{N}_M$ of 
the fraction of molecules that are part of a polymer (inset) and the mean polymer length (main plot)
for various fixed values of scaled coupling. 
In a macroscopic system, $N_M,N \gg1$, two quantities are 
\begin{equation}\label{eq:polydef} 
F_P \doteq\frac{\langle\bar{N}_T\rangle}
{\langle\bar{N}_H\rangle+\langle\bar{N}_T\rangle},\quad 
\langle\bar{L}_P\rangle\doteq F_P \bar{N}_M.
\end{equation}

For zero binding energy the fraction of polymerized molecules increases linearly with $\bar{N}_M$
and the average length of the polymers increases quadratically.
Positive (negative) binding energies enhance (suppress) this rate visibly in both quantities.
Large positive binding energies lures all available molecules into polymerization and the average
length increases linearly.
Large negative binding energy deters all molecules present from polymerization if there is enough
space for monomers.
The average length stays near zero, then rises linearly when polymerization is forced
through crowding.

%
\section{Conclusions}\label{sec:concl}
%
We have explored several themes of fractional exclusion statistics as applied
to particles with shapes.
The context has been Ising spin chains, but the conclusions are readily transcribed to different 
applications.

Consistency in the combinatorics and statistical mechanics of indistinguishable particles at different 
levels of description has been the main theme.
Concerns about this issue were raised previously \cite{Angh10, Wu10}.
Anghel derived a set of necessary conditions that must be satisfied when species of statistically
interacting particles are merged or split \cite{Anghel}.
We have identified two distinct types of mergers that guarantee consistency in the combinatorial 
analysis, both of which confirm Anghel's rules.
Our results confirm that Wu's method of statistical mechanical analysis \cite{Wu94} provides
consistent exact results at each level of description, i.e. before and after mergers.
We have shown evidence that any sufficient conditions for splitting species must be model specific, 
in agreement with Refs~\cite{Anghel, Angh10, Wu10}.

One subsidiary theme has been about open and closed systems of statistically interacting particles.
In the Ising context the numbers of particles from each species fluctuate independently.
We have described two situations where such fluctuations are constrained by a conservation law.
A second subsidiary theme has been about the transcription of motifs for statistically interacting
particles between models for different physical situations.

The statistical mechanics of particles with shapes is a field wide open to applications 
of strong current interest including granular matter jammed in narrow channels
\cite{AB09} and DNA transformed under tension \cite{sarkar, VMRW05}.

\appendix

%
\section{Ising chain with dimerized coupling}\label{sec:dim}
%
The $s=\frac{1}{2}$ Ising chain with alternating bond strengths,
\begin{equation}
  \label{eq:b-1}
 \mathcal{H} =\sum_{l=1}^{2N}\Big(J_{p}-(-1)^{l}J_{m}\Big)
 S_{l}^{z}S_{l+1}^{z},
\end{equation}
where $J_{O}=J_{p}+J_{m}$ is the coupling for bonds with odd $l$ and
$J_{E}=J_{p}-J_{m}$ for bonds with even $l$, has four distinct $T=0$ phases:
\begin{align}\label{eq:dimphases} 
& \Phi_{++}:~ |\uparrow\downarrow\uparrow\downarrow\cdots\rangle_1, 
|\downarrow\uparrow\downarrow\uparrow\cdots\rangle_1 ~\mathrm{at}~ J_O>0, J_E>0;\nonumber \\
& \Phi_{+-}:~ 
|\uparrow\downarrow\downarrow\uparrow\uparrow\downarrow\downarrow\uparrow\cdots\rangle_2
 ~\mathrm{at}~ J_O>0, J_E<0, \nonumber \\
& \Phi_{-+}:~ 
 |\uparrow\uparrow\downarrow\downarrow\uparrow\uparrow\downarrow\downarrow\cdots\rangle_2
 ~\mathrm{at}~ J_O<0, J_E>0, \nonumber \\
& \Phi_{--}:~ |\uparrow\uparrow\cdots\rangle_1, 
|\downarrow\downarrow\cdots\rangle_1 ~\mathrm{at}~ J_O<0, J_E<0.
\end{align}

The periodicity takes the doubling of the unit cell into account.
One set of particles that generates the spectrum of (\ref{eq:b-1}) 
on a lattice with $2N$ sites is closely related with the set of particles from 
Table~\ref{tab:dlj3-specsneel32} for (\ref{eq:1}) on a lattice with $N$ sites.
The Hilbert space of both systems has dimensionality $4^N$.
We transform the inhomogeneous system into a homogeneous system by introducing
compound site variables constructed from pairs of site variables coupled by $J_O$.
The four compound site variables combine to 16 bonds as shown in Table~\ref{tab:cosiva}.
The energy of each compound bond has two parts: one unit of $J_E$-bond energy
(counted fully) and two units of $J_O$-bond energy (each counted half).
Each compound site and each compound bond is also assigned a spin.

\begin{table}[b]
  \caption{Compound sites associated with $J_O$-bonds, their energies,
  and their spin content (top) Compound bonds associated with $J_E$-bonds of
  compound sites, their energies, and their s pin content (bottom).}\label{tab:cosiva} 
\begin{center}
\begin{tabular}{c|cccc}  \rule[-2mm]{0mm}{5mm} 
$J_O$-bond & $\uparrow\uparrow$ & $\downarrow\downarrow$ & $\uparrow\downarrow$ 
& $\downarrow\uparrow$ \\  \hline \rule[-2mm]{0mm}{5mm} 
compound site & $\Uparrow$ & $\Downarrow$ & $\cap$ & $\cup$ \\ \rule[-2mm]{0mm}{5mm}
energy & $+\frac{1}{4}J_O$ & $+\frac{1}{4}J_O$ & $-\frac{1}{4}J_O$ 
& $-\frac{1}{4}J_O$  \\ \rule[-2mm]{0mm}{5mm}
spin & $+\frac{1}{2}$ & $+\frac{1}{2}$ & $0$ & $0$
\end{tabular}

\vspace*{5mm}
\begin{tabular}{ccc|ccc}  \rule[-2mm]{0mm}{5mm} 
$J_E$-bond & energy & spin~ & ~$J_E$-bond & energy & spin \\ \hline \rule[-2mm]{0mm}{5mm} 
$\Uparrow\Uparrow$ &  $~~\frac{1}{4}(J_O+J_E)$ & $+1$ &
$\Uparrow\cap$ & $~~\frac{1}{4}J_E$ & $+\frac{1}{2}$ \\  \rule[-2mm]{0mm}{5mm} 
$\Downarrow\Downarrow$ &  $~~\frac{1}{4}(J_O+J_E)$ & $-1$ &
$\Downarrow\cup$ & $~~\frac{1}{4}J_E$ & $-\frac{1}{2}$ \\  \rule[-2mm]{0mm}{5mm} 
$\Uparrow\Downarrow$ &  $~~\frac{1}{4}(J_O-J_E)$ & $~0$ &
$\cap\Downarrow$ & $~~\frac{1}{4}J_E$ & $-\frac{1}{2}$ \\  \rule[-2mm]{0mm}{5mm} 
$\Downarrow\Uparrow$ &  $~~\frac{1}{4}(J_O-J_E)$ & $~0$ &
$\cup\Uparrow$ & $~~\frac{1}{4}J_E$ & $+\frac{1}{2}$ \\  \rule[-2mm]{0mm}{5mm} 
$\cap\cap$ &  $-\frac{1}{4}(J_O+J_E)$ & $~0$ &
$\cap\Uparrow$ & $-\frac{1}{4}J_E$ & $+\frac{1}{2}$ \\  \rule[-2mm]{0mm}{5mm} 
$\cup\cup$ &  $-\frac{1}{4}(J_O+J_E)$ & $~0$ &
$\cup\Downarrow$ & $-\frac{1}{4}J_E$ & $-\frac{1}{2}$ \\  \rule[-2mm]{0mm}{5mm} 
$\cap\cup$ &  $~~\frac{1}{4}(J_E-J_O)$ & $~0$ &
$\Uparrow\cup$ & $-\frac{1}{4}J_E$ & $+\frac{1}{2}$ \\  \rule[-2mm]{0mm}{5mm} 
$\cup\cap$ &  $~~\frac{1}{4}(J_E-J_O)$ & $~0$ &
$\Downarrow\cap$ & $-\frac{1}{4}J_E$ & $-\frac{1}{2}$ \\  \rule[-2mm]{0mm}{5mm} 
\end{tabular}
\end{center}
\end{table} 

We now select the physical vacuum at $J_O<0$, $J_E>0$, i.e. phase $\Phi_{-+}$, as the 
pseudo-vacuum of a set of particles that generate the spectrum of (\ref{eq:b-1}).
We take advantage of the isomorphism,
\begin{equation}\label{eq:b-2} 
(\Uparrow,~\Downarrow,~\uparrow,~\downarrow) ~\longleftrightarrow~ 
(\Uparrow,~\Downarrow,~\cap,~\cup),
\end{equation}
that transforms the site variables of (\ref{eq:1}) into the compound site 
variables of (\ref{eq:b-1}).
It maps the pseudo-vacuum of the particles listed in Table~\ref{tab:dlj3-specsneel32} 
for (\ref{eq:1}) into the the pseudo-vacuum selected here for (\ref{eq:b-1}).

The particles thus constructed for (\ref{eq:b-1}) have different energies $\epsilon_m$ 
and different spins $s_m$.
The specifications $A_m$, $\alpha_m$, $g_{mm'}$ remain the same.
The energies of particle pairs 15~\&~16, 1~\&~2, 5~\&~6, 3~\&~4 remain equal, making it possible 
to carry out the first four mergers as in Sec.~\ref{sec:spms}.
However, the two motifs of particles 7 and 8 acquire different energies in the transcription,
which appears to make it necessary to split them into subspecies.

Given the indeterminate nature of rules (\ref{eq:angh}) in this process, it requires that the 
combinatorial analysis be performed from scratch for the new subspecies including their 
statistical interactions with all other species.
In the situation at hand, this can be avoided.
Consider the four motifs $\Uparrow\uparrow\Downarrow$, $\Downarrow\uparrow\Uparrow$,
$\Uparrow\downarrow\Downarrow$, $|\Downarrow\downarrow\Uparrow$
of particles 7 and 8.
In the context of (\ref{eq:1}) all four have the same energy at $h=0$ but the first two have spin 
$+\frac{1}{2}$ and the last two spin $-\frac{1}{2}$, producing different Zeeman energies.
In the context of (\ref{eq:b-1}) all four have zero spin but the energy of the first and the last is 
different from the energy of the second and the third.
Can we switch the second and the fourth motif such that 
particles 7 and 8 each have a definite energy again?
The answer is affirmative.
The switch leaves all results of the combinatorial analysis invariant provided the merger of 
species 15~\&~16 into $\bar{15}$ is in place.
We thus end up with 13 species of particles.
Their specifications are listed in Table~\ref{tab:specsalt}.

The statistical mechanical analysis of these particles proceeds analogous to that reported in 
Sec.~\ref{sec:os-d}.
The particles have the same motifs (albeit transcribed and slightly shuffled) but different 
energies.
The same motifs encode different physical structures and produce different kinds of ordering
in the limit $T\to0$.
Adding a Zeeman term to (\ref{eq:a-1}) poses no serious challenge.
It requires that the mergers of 3~\&~4, 5~\&~6, 1~\&~2 be reversed and that the merger
of 15~\&~16 be kept.
The statistical mechanical analysis then proceeds with 16 species and with Zeeman energies 
$-s_mh$ added to $\epsilon_m$.

\begin{widetext}

\begin{table}[htb]
  \caption{Specifications of 13 particles that generate the spectrum of (\ref{eq:b-1}) from 
  the state $(n_{pv}=2)$, 
  \mbox{$|\uparrow\uparrow\downarrow\downarrow\uparrow
  \uparrow\downarrow\downarrow\cdots\rangle_2
   \hat{=} |\Uparrow\Downarrow\Uparrow\cdots\rangle_2$} after four mergers and two splits 
   as described in the text: 
  motif, species, energy (relative to vacuum), spin, capacity constants, and size constants, 
  statistical interaction coefficients. 
  Segments of $\ell$ (compund) vacuum bonds $\Uparrow\Downarrow,\Downarrow\Uparrow$ have energy 
  $E=\frac{1}{4}\ell(J_O-J_E)$.  
  The first species is a compact, the last two are hybrids, and all others hosts.}\label{tab:specsalt} 
\begin{center}
{\begin{tabular}{cc|cccc}
motif& $m$ & $2\epsilon_{m}$ & $s_{m}$ & $A_{m}$ &$\alpha_m$
\\ \hline \rule[-2mm]{0mm}{6mm}
$\Uparrow\Uparrow,\Downarrow\Downarrow$ & $\bar{1}$ & $J_E$ & $\pm1$ & $N-1$ & $1$
\\ \rule[-2mm]{0mm}{5mm}
$\Uparrow\cup\Uparrow,\Downarrow\cap\Downarrow$ & $\bar{3}$ & $J_E-J_O$ & $\pm1$ & 
$N-2$ & $2$ \\ \rule[-2mm]{0mm}{5mm}
$\Uparrow\cap\Uparrow,\Downarrow\cup\Downarrow$ & $\bar{5}$ & $J_E-J_O$ & 
$\pm1$ & $N-2$ & $2$ \\ \rule[-2mm]{0mm}{5mm}
$\Uparrow\cap\Downarrow,\Downarrow\cup\Uparrow$ & $7$ &
$2J_E-J_O$ & $0$ & $N-2$ & $2$ \\ \rule[-2mm]{0mm}{5mm}
$\Downarrow\cap\Uparrow,\Uparrow\cup\Downarrow$ & $8$ & 
$-J_O$ & $0$ & $N-2$ & $2$ \\ \rule[-2mm]{0mm}{5mm}
$\Uparrow\cap\cup\Uparrow$ & $9$ & 
$3J_E-2J_O$ & $+1$ & $\frac{N-3}{2}$ & $3$ \\ \rule[-2mm]{0mm}{5mm}
$\Downarrow\cap\cup\Downarrow$ & $10$ & 
$J_E-2J_O$ & $-1$ & $\frac{N-3}{2}$ & $3$ \\ \rule[-2mm]{0mm}{5mm}
$\Uparrow\cup\cap\Uparrow$ & $11$ & 
$J_E-2J_O$ & $+1$ & $\frac{N-3}{2}$ & $3$ \\ \rule[-2mm]{0mm}{5mm}
$\Downarrow\cup\cap\Downarrow$ & $12$ & 
$3J_E-2J_O$ & $-1$ & $\frac{N-3}{2}$ & $3$ \\ \rule[-2mm]{0mm}{5mm}
$\Uparrow\cup\cap\Downarrow,\Downarrow\cap\cup\Uparrow$ & $13$ & 
$2J_E-2J_O$ & $0$ & $N-3$ & $3$ \\ \rule[-2mm]{0mm}{5mm}
$\Uparrow\cap\cup\Downarrow,\Downarrow\cup\cap\Uparrow$ & $14$ & 
$2J_E-2J_O$ & $0$ & $N-3$ & $3$ \\ \rule[-2mm]{0mm}{5mm}
$\cap\cap,\cup\cup$ & $\bar{15}$ & $-J_O$ & $0$ & 
$0$ & $1$ \\ \rule[-2mm]{0mm}{5mm}
$\cap\cup\cap,\cup\cap\cup$ & $17$ & $2J_E-2J_O$ & $0$ & $0$ & $2$
\end{tabular}\hspace{9mm}%
\begin{tabular}{c|rrrrrrrrrrrrr} 
$g_{mm'}$ & $~\bar{1}$ & $\bar{3}$ & $\bar{5}$ & $7$ & $8$ & $9$ & 
$10$ & $11$ & $12$ & $13$ & $14$ & $\bar{15}$ & $17$ \\
\hline \rule[-2mm]{0mm}{6mm}
$\bar{1}$ & $1$ & $1$ & $1$ & $1$ & $1$ & $2$ 
& $2$ & $2$ & $2$ & $2$ & $2$ & $1$ & $2$  \\ \rule[-2mm]{0mm}{5mm}
$\bar{3}$ & $1$ & $2$ & $2$ & $1$ & $1$ & $2$ 
& $2$ & $2$ & $2$ & $2$ & $2$ & $1$ & $2$  \\ \rule[-2mm]{0mm}{5mm}
$\bar{5}$ & $1$ & $1$ & $2$ & $1$ & $1$ & $2$ 
& $2$ & $2$ & $2$ & $2$ & $2$ & $1$ & $2$ \\ \rule[-2mm]{0mm}{5mm}
$7$ & $1$ & $2$ & $2$ & $2$ 
& $1$ & $3$ 
& $3$ & $3$ & $3$ & $3$ 
& $3$ & $1$ & $2$ \\ \rule[-2mm]{0mm}{5mm}
$8$ & $1$ & $2$ & $2$ & $2$ 
& $2$ & $3$ 
& $3$ & $3$ & $3$ & $3$ 
& $3$ & $1$ & $2$ \\ \rule[-2mm]{0mm}{5mm}
$9$ & $\frac{1}{2}$ & $1$ & $1$ & $\frac{1}{2}$ 
& $\frac{1}{2}$ & $\frac{3}{2}$ 
& $\frac{3}{2}$ & $\frac{1}{2}$ & $\frac{3}{2}$ & $1$ 
& $1$ & $\frac{1}{2}$ & $1$ \\ \rule[-2mm]{0mm}{5mm}
$10$ & $\frac{1}{2}$ & $1$ & $1$ & $\frac{1}{2}$ 
& $\frac{1}{2}$ & $\frac{3}{2}$ 
& $\frac{3}{2}$ & $\frac{3}{2}$ & $\frac{1}{2}$ & $1$ 
& $1$ & $\frac{1}{2}$ & $1$ \\ \rule[-2mm]{0mm}{5mm}
$11$ & $\frac{1}{2}$ & $1$ & $1$ & $\frac{1}{2}$ 
& $\frac{1}{2}$ & $\frac{3}{2}$ 
& $\frac{3}{2}$ & $\frac{3}{2}$ & $\frac{3}{2}$ & $1$ 
& $1$ & $\frac{1}{2}$ & $1$ \\ \rule[-2mm]{0mm}{5mm}
$12$ & $\frac{1}{2}$ & $1$ & $1$ & $\frac{1}{2}$ 
& $\frac{1}{2}$ & $\frac{3}{2}$ 
& $\frac{3}{2}$ & $\frac{3}{2}$ & $\frac{3}{2}$ & $1$ 
& $1$ & $\frac{1}{2}$ & $1$ \\ \rule[-2mm]{0mm}{5mm}
$13$ & $1$ & $2$ & $2$ & $1$ 
& $1$ & $3$ 
& $3$ & $3$ & $3$ & $3$ 
& $2$ & $1$ & $2$ \\ \rule[-2mm]{0mm}{5mm}
$14$ & $1$ & $2$ & $2$ & $1$ 
& $1$ & $3$ 
& $3$ & $3$ & $3$ & $3$ 
& $3$ & $1$ & $2$ \\ \rule[-2mm]{0mm}{5mm}
$\bar{15}$ & $0$ & $-1$ & $-1$ & $-1$ 
& $-1$ & $-2$ 
& $-2$ & $-2$ & $-2$ & $-2$ 
& $-2$ & $0$ & $-2$ \\ \rule[-2mm]{0mm}{5mm}
$17$ & $0$ & $-1$ & $-1$ & $-1$ 
& $-1$ & $-1$ 
& $-1$ & $-1$ & $-1$ & $-1$ 
& $-1$ & $0$ & $0$
\end{tabular}}
\end{center}
\end{table} 

\end{widetext}

%
%



\begin{thebibliography}{100}

\bibitem{Nolt09}
W. Nolting,
\emph{Fundamentals of Many-Body Physics}
(Springer-Verlag, Berlin, 2009).

\bibitem{Reic98}
L. E. Reichl,
\emph{A Modern Course in Statistical Physics}
(Wiley, New York, 1998).

\bibitem{copic}
D. Liu, P. Lu, G. M\"uller, and M. Karbach,
Phys. Rev. E \textbf{84}, 021136 (2011).

\bibitem{picnnn}
P. Lu, D. Liu, G. M\"uller, and M. Karbach
(unpublished) [arXiv:1108.2990].

\bibitem{CJSZ03}
X. Y. Chen, Q. Jiang, W. Z. Shen, and C. G. Zhong,
J. Magnetism Magnet. Mater. \textbf{262}, 258 (2003).

\bibitem{LVP+08}
P. Lu, J. Vanasse, C. Piecuch, M. Karbach, and G. M{\"u}ller, J. Phys. A \textbf{41}, 265003 (2008).

\bibitem{Hald91a}
F. D. M. Haldane, Phys. Rev. Lett. \textbf{67}, 937 (1991).

\bibitem{Khar05}
A. Khare,
\emph{Fractional Statistics and Quantum Theory}
(World Scientific, Singapore, 2005).

\bibitem{Anghel}
D.-V. Anghel, J. Phys. A \textbf{40}, F1013 (2007);
Europhys. Lett. \textbf{87}, 60009 (2009).

\bibitem{Angh10}
D.-V. Anghel, Phys. Rev. Lett. \textbf{104}, 198901 (2010).

\bibitem{Wu10}
Y.-S. Wu, Phys. Rev. Lett. \textbf{104}, 198902 (2010).

\bibitem{note1}
The relative spin orientation of successive solitons is encoded in the fermionic motif
by the number of holes between particles.
For aligned (anti-aligned) soliton spins that number is even (odd).

\bibitem{note3}
These distinguishable traits may very well become relevant in different contexts.
It will then be necessary to split some or all of the species with multiple motifs into one-motif species.

\bibitem{Wu94}
Y.-S. Wu, Phys. Rev. Lett. \textbf{73}, 922 (1994).

\bibitem{note4}
A more systematic study of entropy landscapes is found in Ref.~\cite{picnnn} for a different model.

\bibitem{note5}
The model can be adjusted to accommodate different distance ratios.

\bibitem{AB09}
R. K. Bowles and I. Saika-Voivod,
Phys. Rev. E \textbf{73}, 011503 (2006);
S. S. Ashwin and R. K. Bowles,
Phys. Rev. Lett. \textbf{102}, 235701 (2009).

\bibitem{sarkar}
A. Sarkar,  J. F. L\'eger,  D. Chatenay,  and J. F. Marko,
Phys. Rev. E \textbf{63}, 051903 (2001);
 J.-F. L\'eger, G. Romano,  A. Sarkar, J. Robert , L. Bourdieu, D. Chatenay, and J. F. Marko,
Phys. Rev. Lett. \textbf{83}, 1066 (1999).

\bibitem{VMRW05}
I. D. Vladescu, M. J. McCauley, I. Rouzina, and M. C. Williams,
Phys. Rev. Lett. \textbf{95}, 158102 (2005);
J. van Mameren, Proc. Natl. Acad. Sci. USA \textbf{106}, 18231 (2009).

\end{thebibliography}
\end{document}